\documentclass[fleqn,usenatbib]{mnras}

\usepackage{newtxtext,newtxmath}
\usepackage[T1]{fontenc}
\DeclareRobustCommand{\VAN}[3]{#2}
\let\VANthebibliography\thebibliography
\def\thebibliography{\DeclareRobustCommand{\VAN}[3]{##3}\VANthebibliography}
\usepackage{graphicx}	
\usepackage{amsmath}	
\usepackage{xcolor}     
\usepackage{makecell}
\usepackage{hyperref}
\usepackage{verbatim}
\usepackage{comment}
\usepackage{ulem}
\usepackage{lineno}
\usepackage{soul}

\newcommand{\desimtwo}{DESI-M2}
\newcommand{\ihMpc}{\,h^{-1}{\rm Mpc}}
\newcommand{\ihGpc}{\,h^{-1}{\rm Gpc}}

\newcommand{\rascalc}{{\tt RascalC}}

\title[DESI Early BAO Detection]{First Detection of the BAO Signal from Early DESI Data}

\author[J. Moon et al. (2023)]{\parbox{\textwidth}{
Jeongin Moon,$^{1}$\thanks{E-mail: \url{graziano@sejong.ac.kr} (GR); \url{jmoon@mpe.mpg.de} (JM)} 
David Valcin,$^{2}$
Michael Rashkovetskyi,$^{3}$
Christoph Saulder,$^{4}$
Jessica Nicole Aguilar,$^{5}$ 
Steven Ahlen,$^{6}$  
Shadab Alam,$^{7}$  
Stephen Bailey,$^{5}$ 
Charles Baltay,$^{8}$ 
Robert Blum,$^{9}$ 
David Brooks,$^{10}$ 
Etienne Burtin,$^{11}$ 
Edmond Chaussidon,$^{11}$
Kyle Dawson,$^{12}$
Axel de la Macorra,$^{13}$
Arnaud de Mattia,$^{11}$
Govinda Dhungana,$^{14}$
Daniel Eisenstein,$^{3}$
Brenna Flaugher,$^{15}$
Andreu Font-Ribera,$^{16}$
Jaime E. Forero-Romero,$^{17}$
Cristhian Garcia-Quintero,$^{18}$
Satya Gontcho A Gontcho,$^{5}$
Julien Guy,$^{5}$
Malik Muhammad Sikandar Hanif,$^{19}$
Klaus Honscheid,$^{20,21}$
Mustapha Ishak,$^{18}$
Robert Kehoe,$^{22}$
Sumi Kim,$^{23}$
Theodore Kisner,$^{5}$
Anthony Kremin,$^{5}$
Martin Landriau,$^{5}$
Laurent Le Guillou,$^{24}$
Michael Levi,$^{5}$
Marc Manera,$^{25}$
Paul Martini,$^{26,21}$
Patrick McDonald,$^{5}$
Aaron Meisner,$^{9}$
Ramon Miquel,$^{16,27}$
John Moustakas,$^{28}$
Adam Myers$,^{29}$
Seshadri Nadathur,$^{30}$
Richard Neveux,$^{31}$
Jeffrey A. Newman,$^{32}$
Jundan Nie,$^{33}$
Nikhil Padmanabhan,$^{8}$
Nathalie Palanque-Delabrouille,$^{5,11}$
Will Percival,$^{34,35}$
Alejandro P{\'e}rez Fern{\'a}ndez,$^{13}$
Claire Poppett,$^{36,5}$
Francisco Prada,$^{37}$
Anand Raichoor,$^{5}$
Ashley J. Ross,$^{26,21}$
Graziano Rossi,$^{1{\color{blue} \star}}$  
Lado Samushia,$^{38}$
David Schlegel,$^{5}$
Hee-Jong Seo,$^{2}$
Gregory Tarl{\'e},$^{19}$
Mariana Vargas Magana,$^{13}$
Andrei Variu,$^{39}$
Benjamin Alan Weaver,$^{9}$
Martin J. White,$^{40}$
Christophe Y{\`e}che,$^{11}$
Sihan Yuan,$^{41}$
Cheng Zhao,$^{39}$
Rongpu Zhou,$^{5}$
Zhimin Zhou,$^{33}$ and
Hu Zou$^{33}$
} \vspace*{6pt} \\ 
Affiliations are listed at the end of the paper}

\date{Accepted 2023 August 29. Received 2023 August 28; in original form 2023 April 19}
\pubyear{2023}

\begin{document}
\label{firstpage}
\pagerange{\pageref{firstpage}--\pageref{lastpage}}
\maketitle

%%%%%%%%%%%%%%%%%%%%%%%%%%%%%%%%%%%%%%%%%%%%%%%%%%
%%%%%%%%%%%%%%%%%%%%%%%%%%%%%%%%%%%%%%%%%%%%%%%%%%
%%%  ABSTRACT

\begin{abstract}
We present the first detection of the baryon acoustic oscillations (BAO) 
signal obtained using unblinded data collected during the initial two months of 
operations of the Stage-IV ground-based Dark Energy Spectroscopic Instrument (DESI).
From a selected sample of $261,291$ Luminous Red Galaxies  
spanning the redshift interval $0.4 < z < 1.1$ and covering $1651$ square degrees 
with a $57.9\%$ completeness level, we report
a $\sim 5\sigma$ level BAO detection and the measurement of the BAO location at 
a precision of $1.7\%$. Using a  Bright Galaxy Sample of $109,523$ galaxies 
in the redshift range $0.1 < z < 0.5$, 
over $3677$ square degrees with a $50.0\%$ completeness,
we also detect the BAO feature at $\sim 3\sigma$ significance with a $2.6\%$ precision. 
These first BAO measurements represent an important milestone, 
acting as a quality control on the optimal performance of the
complex robotically-actuated, fiber-fed DESI spectrograph,
as well as an early validation of the DESI
spectroscopic pipeline and data management system.  
Based on these first promising results, we forecast that DESI is
on target to achieve
a high-significance BAO detection at sub-percent precision
with the completed 5-year survey data,
meeting the
top-level science requirements on BAO measurements.  
This exquisite level of precision will set new standards in cosmology and confirm DESI as the most competitive
BAO experiment for the remainder of this decade.
\end{abstract}

%%%%%%%%%%%%%%%%%%%%%%%%%%%%%%%%%%%%%%%%%%%%%%%%%%
%%%%%%%%%%%%%%%%%%%%%%%%%%%%%%%%%%%%%%%%%%%%%%%%%%
%%%  KEYWORDS

\begin{keywords}
cosmology: large-scale structure of Universe, observations, dark energy -- galaxies: statistics -- methods: data analysis, statistical.
\end{keywords}

%%%%%%%%%%%%%%%%%%%%%%%%%%%%%%%%%%%%%%%%%%%%%%%%%%
%%%%%%%%%%%%%%%%%%%%%%%%%%%%%%%%%%%%%%%%%%%%%%%%%%
%%%  INTRODUCTION

\section{Introduction}

%--------------------------------------------------------------- 
%---------------------------------------------------------------  

The precise measurement of the expansion history of the Universe remains one 
of the key challenges in modern cosmology, and represents a
compelling probe of the nature of dark energy (DE).
The distance-redshift relation over a wide redshift range tests whether the 
accelerated expansion is consistent with a 
cosmological constant ($\Lambda$) or requires a dynamical explanation. It is also an important constraint
on the growth rate of structures, allowing precise probes of gravity on 
cosmological scales, and on the Hubble constant, shedding light on the source of 
the ``Hubble tension'' as coming either from as-yet unappreciated astrophysical 
systematics or new physics. Finally, it breaks cosmological parameter degeneracies 
in, e.g., neutrino mass measurements. 
Recent results from state-of-the-art experiments 
have provided highly accurate constraints on the basic parameters 
of the standard spatially flat $\Lambda$CDM cosmological model, dominated by collisionless cold dark matter (CDM) 
and a  DE component in the form of $\Lambda$ \citep{Planck2020cosmo,SDSS-DR16-cosmology,DESCosmology}.

The baryon acoustic oscillation (BAO) method is one of the most mature and robust probes of
expansion history. Acoustic oscillations in the early pre-recombination Universe imprint a 
feature in the galaxy distribution 
at a scale ($r_{\rm d}$) set by the sound horizon evaluated at the drag epoch. 
The physics of these oscillations and 
the scale of this feature (which constitutes a fundamental standard ruler) 
are exquisitely calibrated by cosmic microwave background (CMB)
measurements. Furthermore, the scale of the sound horizon is  
much bigger than the scale of physics of 
nonlinear structure formation and galaxy biasing, making it robust to the subsequent evolution of the Universe;
for a review on BAO, see \cite{BAOreview2013} and references therein.
The apparent size of this standard ruler across and along the line of sight characterizes the angular
diameter distance ($D_{\rm A}$) and the Hubble parameter ($H$) as a function of redshift.
Previous surveys have successfully measured these quantities directly from the BAO 
feature at different redshifts. Examples of first BAO detections obtained from multiple tracers include 
\cite{BAO-discovery}, \cite{BAO2dF}, \cite{Wigglez}, \cite{LyABAO},
while the most recent results are reported in \cite{SDSS-DR16-cosmology} and in \cite{DESCosmology}.
 
BAO measurements at sub-percent precision are considered
primary science targets for the Dark Energy Spectroscopic Instrument \citep[DESI;][]{DESICollaboration2016a},
along with novel constraints on theories of modified
gravity and inflation, and on neutrino masses. 
DESI, the only Stage-IV DE experiment that is currently taking data, 
aims to provide multiple sub-percent 
distance measurements 
over a broad $0 < z < 3.5$  
redshift range. DESI represents an 
order-of-magnitude improvement both in the volume surveyed and in the number of galaxies measured
over previous experiments -- e.g., 
Extended Baryon Oscillation Spectroscopic Survey
\citep[eBOSS;][]{Dawson2016},
a key component of the  fourth generation \citep[SDSS-IV;][]{Blanton2017}
of the Sloan Digital Sky Survey 
\citep[SDSS;][]{York2000}. 
In addition, DESI builds in a number of internal systematics checks using
multiple tracer populations to probe common volumes.

Given the exquisite precision achievable by the DESI survey, the DESI collaboration decided to 
blind the redshift data to avoid any confirmation biases that can potentially impact all of the cosmological analyses. 
A general procedure to blind a modern redshift survey has been discussed in \citet{brieden2020}, and the 
exact implementation into the DESI framework will be described elsewhere. However, for
early quality assurance tests and in order to validate the data processing and analysis pipelines,   
the first two months of DESI observations (hereafter referred to as \desimtwo) have been kept 
intentionally unblinded. 

In this work, we use the \desimtwo\ dataset and 
report the first high-significance detection of the BAO signal 
from the initial two months of DESI operations.
As part of testing these early data, we 
integrate the eBOSS BAO pipeline into the DESI analysis framework, and 
apply such pipeline to measure the BAO scale with   
updates to accommodate all of the DESI specifics. 
While the DESI survey has four primary galaxy tracer 
populations to measure clustering,\footnote{The DESI survey also has a fifth tracer, i.e., the Lyman-$\alpha$ (Ly$\alpha$) forest.} the survey
strategy implies that not all tracers will have the same completeness in 
the very early data. The 
two most complete samples are the DESI Bright Galaxy Sample (BGS) and the Luminous Red Galaxies 
(LRGs). We focus on these two samples here for the BAO measurement, since our simulations 
suggest that we would not expect a BAO detection in the Emission Line Galaxy (ELG) and Quasar
(QSO) samples given the number density, the low completeness, and the volume of this early data.
A similar signal-to-noise (SN) consideration led us to concentrate on the isotropic distance measurements 
$\alpha \equiv D_{\rm A}^{2/3} H^{-1/3}/r_{\rm d}$, probed by the angle-averaged galaxy correlation function (the "monopole"). 
Future DESI analyses will present measurements using all four tracer populations, as well as 
measurements of $D_{\rm A} H$ from
the Alcock-Paczynski effect \citep{Alcock1979}.

As we will show in our analysis, even these early data yield 
a precision in distance comparable to measurements from previously completed surveys 
\citep[i.e.,][]{BOSSDR9BAO,Bautista2021}, 
highlighting the remarkable statistical power of the DESI data. 
In the spirit of the DESI blinding policy, we restrict ourselves
to providing just the statistical precision of the measurements rather than the actual distance 
values -- which will be presented instead in a series of DESI Year 1 (Y1) forthcoming 
cosmological papers. This work therefore should be seen as an end-to-end quality assurance  
of the DESI data management system, as well as an early validation of the DESI spectroscopic pipeline.  

%---------------------------------------------------------------  

% Paper Layout
 
The layout of the paper is organized as follows. In Section 
\ref{sec:datasamples}, we briefly describe the main aspects of the
\desimtwo\ sample used in this work, along with the 
procedure to build the corresponding large-scale structure (LSS) data catalogs. 
In Section \ref{sec:mocks}, we present the approximate and $N$-body-based  
mocks adopted in the core analysis, and explain how 
such synthetic catalogs are constructed in order to mimic the complex 
footprint and characteristics of the \desimtwo.
In Section \ref{sec:methods}, we illustrate all of the analysis tools, namely 
the chosen two-point clustering estimator, the density field reconstruction technique, 
and the BAO fitting methodology. 
Section \ref{sec:cov} addresses covariance matrices, 
and in particular the construction, calibration, and validation on mock data
of semi-analytical semi-empirical covariances for the BAO fitting procedure.
More details on the 
covariance matrices adopted here are reported in a companion paper \citep{RascalC-DA02}.
The main results are detailed in Section \ref{sec:results}, where we
assess the precision and detection statistics of the BAO feature in the LRG and BGS samples. 
We then briefly address the expected precision of the final Year 5 (Y5) DESI LRG sample in Section \ref{sec:outlook}, 
in terms of the BAO detection level, based on forecasts obtained from our promising early results.
Finally, we conclude in Section \ref{sec:conclusion}, where we summarize the main findings 
and highlight the relevance for the upcoming Y1 DESI dataset.
We also leave some additional
material in Appendix \ref{sec:appendix}. 

%%%%%%%%%%%%%%%%%%%%%%%%%%%%%%%%%%%%%%%%%%%%%%%%%%
%%%%%%%%%%%%%%%%%%%%%%%%%%%%%%%%%%%%%%%%%%%%%%%%%%
%%%  DATA

\section{DESI Main Survey Data: First Two Months}\label{sec:datasamples}

%--------------------------------------------------------------- 
%---------------------------------------------------------------  

In this section, we provide a concise description of the \desimtwo\ dataset, along with
several specifics on the LSS catalog construction. A number of additional technical details 
can be found in the quoted supporting papers, many of which are still in a preparatory phase and will be
available at the time of the official Y1 DESI data release.
 
%--------------------------------------------------------------- 
%---------------------------------------------------------------  

\subsection{DESI Early Data: General Aspects} 
 
 %-------------------------------%

\begin{table*}
\centering
\caption{Statistics of the four primary DESI targets from the \desimtwo\ dataset, including completeness information.}
\begin{tabular}{c|cccccc}
\hline\hline
Target & $N_{\rm North}$ & $N_{\rm South}$ & $N_{\rm Total}$ & $z$ range & Area [deg$^2$] & Completeness  \\\hline
BGS Bright & 239492 & 390988 & 630480 & 0.1 - 0.5 & 3677 & 0.500\\
BGS Bright, $M_r<-21.5$ & 38472 & 71051 & 109523 & 0.1 - 0.5 & 3677 & 0.500\\
LRG  & 80651 & 180640 & 261291 & 0.4 -1.1 & 1651 & 0.579\\
ELG  & 55383 & 117145 & 172528 & 0.8 - 1.6 & 976 & 0.297\\
QSO & 70337 & 153453 & 223790 & 0.8 - 3.5 & 2906 & 0.778\\
\hline\hline
\end{tabular}
\label{tab:edatargets}
\end{table*}

 %-------------------------------%

DESI began its main program on May 17, 2021. Its commissioning and `Survey Validation' (SV) phases \citep{DESIsv} had proved
the instrument \citep{DESI_instrument} and operations strategy \citep{DESIsops} to be efficient.
The DESI collaboration decided that the first two months of the observations of DESI main survey data (i.e., \desimtwo) could be analyzed without the 
blinding restrictions imposed on the rest of the sample that will be used for DESI Y1 Key Projects.

The \desimtwo\ data were observed on nights in 2021 from May 14th through July 9th on 304 dark time and 342 bright time `tiles'. 
Each tile represents a specific sky location pointing of the telescope and specific target selection for each of the 5000 robotic positioners 
populating the DESI focal plane \citep{DESIfocalplane} determined by the DESI \texttt{fiberassign} software  (Raichoor et al. in preparation).
The spectra extracted from these observations were reduced by the DESI spectroscopic pipeline \citep{DESIspec} and released to the 
DESI collaboration as the \texttt{Guadalupe} spectroscopic product. The redshift measurements in these \texttt{Guadalupe} data are used in this paper 
and will be made public with the DESI Y1 data release (DR1), i.e., they are not available in the 
DESI early data release (EDR; \citealt{EDR}).\footnote{The analogous spectroscopic data reductions and redshift fits for the EDR are \texttt{Fuji}
and will be publicly available on NERSC here: \url{https://data.desi.lbl.gov/public/edr/spectro/redux/fuji}}  

The DESI-M2 tiles are primarily first pass tiles that do not overlap each other. In dark time, 
the full DESI survey observes tiles in 7 
overlapping passes, with a median overlap of 5 \citep{DESIsops}. Thus, the DESI-M2 data are substantially 
less complete in the area observed than they will be when the survey is finished. This incompleteness affects all samples, 
but is most extreme for the target classes that are given the lowest priority during the assignment of fibers on a tile 
(DESI fiber assignment is reported in Raichoor et al. in preparation). We describe this further next, 
when discussing the different DESI target classes.

%--------------------------------------------------------------- 
%---------------------------------------------------------------  

\subsection{DESI Targets}\label{subsec:targets}

DESI divides its observing time into a `bright' and a `dark' time program, for which the targeting is done independently \citep{DESItarget}. 
During dark time, in order of priority for fiber assignment, 
QSOs \citep{DESIqsotarget}, LRGs \citep{DESIlrgtarget}, and ELGs \citep{DESIelgtarget} are observed. QSOs with redshifts greater 
than $2.1$ are selected for follow-up in order to increase the SN of the spectra in the Ly$\alpha$ 
forest region. During bright time, a BGS \citep{DESIbgstarget} is observed, which has a `Bright' and `Faint' 
component, as well as Milky Way stars (MWS; \citealt{DESI_target_MWS}). In this work, we only consider 
the higher priority BGS Bright sample.  

\begin{figure}
    \centering 
    \includegraphics[width=1.01\columnwidth]{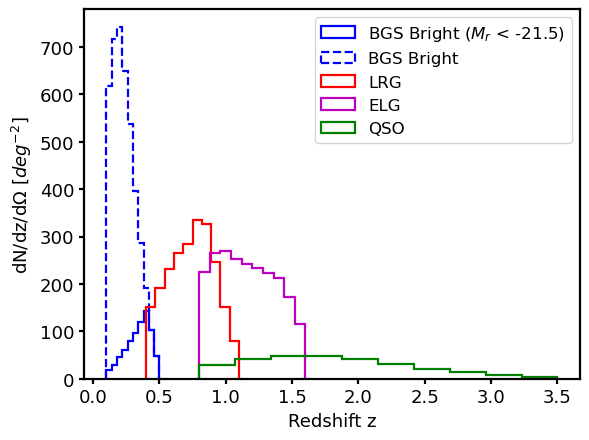}
    \caption{Redshift distribution of the four primary DESI tracers, from the \desimtwo\ clustering catalogs.}
    \label{fig:dndz}
\end{figure}

For detailed discussions of how these target samples were chosen, we refer the reader to the individual 
selection papers previously cited. 
Table \ref{tab:edatargets} summarizes key properties of the samples and Figure \ref{fig:dndz} shows the redshift distribution of each sample. 
Combined, they will allow measurements of large-scale clustering modes at better than the sample variance limit to $z<1.6$ \citep{DESIsv}. 
The QSO sample provides this information at a lower sampling rate all the way to redshifts greater than $3$ and further samples density fluctuations 
via the variance of Ly$\alpha$ forest absorption in each spectrum. The BGS sample is approximately flux limited and thus has 
a spatial density that rapidly increases as the redshift gets lower and is approximately sample variance limited to $z<0.5$.  
There is also substantial overlap between the LRG and ELG catalogs, which will allow comparison between results obtained from the most massive and 
passive galaxies (LRG) and those that are actively star forming (ELG).

%-------------------------------%

%-------------------------------%

\begin{figure}
    \centering
    \includegraphics[width=1.0\columnwidth]{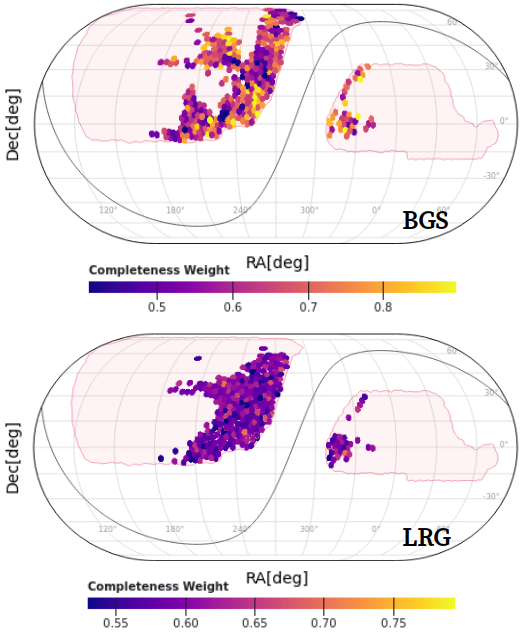}
    \caption{Footprints of the \desimtwo\ BGS Bright (top) and LRG (bottom) clustering samples, color-coded by completeness weights.  
    The total areas highlighted by the pink color represent the final DESI Y5 expected footprint. 
    The specific \desimtwo\ areas covered by the BGS Bright and LRG samples are respectively 3677 deg$^2$ and 1651 deg$^2$.}
    \label{fig:footprint}
\end{figure}    

%-------------------------------%

%--------------------------------------------------------------- 
%---------------------------------------------------------------  

\subsection{LSS Catalog Construction}
\label{sec:LSScat}

The construction of the LSS catalogs involves determining the area on the sky where good observations were possible for each tracer, applying criteria 
on the DESI data to select reliable redshifts within a given redshift range, and providing weights that correct for variations in observing completeness, 
target density due to changes in imaging conditions, and relative redshift success due to variations in DESI observing. 
The overall process is similar to that applied to SDSS (most recently eBOSS; \citealt{ebossLSS}), with the specific details of 
DESI observations accounted for as we describe here. The pipeline that was applied to the \desimtwo\ sample represents an 
early version of the DESI LSS catalog pipeline, which will be fully described (and considerably improved) in Y1 publications. 
Many aspects of the pipeline match that applied to the 
DESI `One Percent Survey',   
which is detailed in the overall description of the DESI EDR \citep{EDR}. In what follows, we provide details on the specific choices applied for \desimtwo.

%---------------------------------------------------------------  

% Randoms
 
The `randoms' that populate the sky area where good observations were possible were produced using 
the same procedures as applied to 
the DESI One Percent Survey LSS catalogs. DESI randoms are produced using a 
standard such that each individual (and independent) set has a density of $2500$ deg$^{-2}$. We use $10$ such sets for the 
\desimtwo\ clustering measurements and thus the total sky density of the random samples used is $25000$ deg$^{-2}$. 
The process of creating DESI randoms produces significantly different areas for different tracer types due to the priority masking 
(e.g., we have no randoms in areas where LRGs could not have been observed because a higher priority QSO target 
was assigned to the fiber positioner associated with that sky location). 
In order to determine the effective area occupied by each sample, we simply count the number of randoms in one of the 
final LSS random catalogs sets and divide by $2500$ deg$^{-2}$. 
These areas are given in Table \ref{tab:edatargets}. While the total effective area is considerably different per 
tracer, the footprint of tiles is the same for all 
dark/bright time tracers. Figure \ref{fig:footprint} shows the footprint of tiles for 
BGS (bright time; top panel) and LRG (dark time; bottom panel) 
tracers. The plot is constructed via a web interface 
provided by David Kirkby,\footnote{\url{https://observablehq.com/@dkirkby/skymap/}} 
and it is color coded by the completeness in each tile grouping.

%---------------------------------------------------------------  

% Data

For the data samples, we follow the same procedures as applied to the One Percent Survey LSS catalogs (described with the EDR) 
in order to select targets of the given type that could have been observed. Any unobserved targets at this stage were not observed because 
a target of the same type was instead observed at the given sky location. Each observed target is given a  completeness weight, WEIGHT\_COMP, 
equal to the total number of targets (of the given type) at the location of the observed target (with unique `locations' determined by the combination observed tile and fiber positioner; see \citealt{EDR} for more details). In particular, we note that the fiber patrol radius is at most 89'': it depends on e.g., the focal plane position due to the optics.

The criteria for the BGS and LRG samples we focus this study on are:
\begin{itemize}
    \item LRG: $0.4 < z < 1.1$, ZWARN$=$0, DELTACHI2$>$15 
    \item BGS: $0.1 < z < 0.5$, ZWARN$=$0, DELTACHI2$>$40. 
\end{itemize}
ZWARN is a bitmask generated by the redshift pipeline \citep{DESIspec}, where any non-zero value indicates a problem. The criteria on 
DELTACHI2, which is the difference in $\chi^2$ between the two best-fitting redshift solutions, were shown to provide pure and complete samples in the respective targeting papers (\citealt{DESIbgstarget,DESIlrgtarget}). 
For LRGs, the choice of redshift range is motivated by the fact that the number density is approximately constant at $5\times10^{-4}(h^{-1}{\rm Mpc})^3$ between 
$0.4 < z < 0.8$. At $z>0.8$,  the LRG density decreases mostly due to the sample's minimum flux threshold and is less than $1\times10^{-4}(h^{-1}{\rm Mpc})^3$ for $z>1.1$. 
Similarly, the density of the BGS sample decreases to less than $1\times10^{-4}(h^{-1}{\rm Mpc})^3$ for $z>0.5$. For the BGS sample, 
we also apply an absolute magnitude cut in the $r$-band $M_{\rm r}<-21.5$. When obtaining the absolute magnitude, 
we simply apply the distance modulus, i.e., we do not apply any corrections for evolution (`e' correction) or the shape of the spectrum (`k' correction). 
The cut provides a sample with roughly constant number density at $\sim 8\times 10^{-4}h^{3}$Mpc$^{-3}$ and clustering amplitude for $z<0.4$ and is thus sufficient for our preliminary study. 
Future DESI studies will likely include k+e corrections, especially for the selection of BGS samples.

We then add two more weights in order to account for variations in the selection of the data. 
The first corrects for fluctuations in the 
target data that are due to variation in the imaging data quality. To do so, we apply the  random forest regression 
method \citep{ChaussidonReg} available as an option in the \texttt{regressis} package,\footnote{\url{https://github.com/echaussidon/regressis/releases/tag/1.0.0}}
given maps of imaging properties compiled by the DESI targeting team. The data (after redshift cuts) and randoms are combined to produce a map of 
the projected density of the sample at \texttt{Healpix} \citep{HEALPix} $N_{\rm side}=256$ and is compared to maps of the depth and 
PSF size in the $g$, $r$, $z$, and $W1$ bands, the $E(B-V)$ Galactic extinction according the \cite{SFD} dust maps, and the 
stellar density observed in the Gaia 2nd data release \citep{GaiaDR2}. The \texttt{regressis} random forest method is used to 
determine a model of the projected density fluctuations as a function of those map quantities and the inverse of the model is included in 
the catalogs as a weight, `WEIGHT\_SYS'. For our LRG and BGS samples, very similar clustering results are 
obtained when instead obtaining the weights using the linear regression method applied to eBOSS, described in \citep{ebossLSS}.

After, we obtain a weight to account for variations in redshift success based on the particulars of DESI observations. 
\cite{DESIlrgtarget} showed that the LRG redshift success can be modeled as a function of the effective observing time and the target's fiber flux in the $z$-band. 
A similar dependency exists for BGS, with the $r$-band fiber flux the relevant photometric quantity. The inverse of the best-fit model for the failure rate is used as `WEIGHT\_ZFAIL'. 
We find that applying these redshift failure weights has very little impact on the clustering measurements used in this work.

Next, we determine `FKP' weights \citep{FKP} in order to properly weight each volume element with respect to how 
each sample's number density changes with redshift. This is simply
given by: 
\begin{equation}
    w_{\rm FKP} = \frac{1}{1+n(z)  C  P_0}
\end{equation}
where $n(z)$ is the weighted number per volume, $C$ is the mean completeness for the sample, and $P_0$ is a fiducial power-spectrum amplitude. 
We use $P_0=10^4(h^{-1}{\rm Mpc})^3$ for LRGs and $P_0=7\times10^3(h^{-1}{\rm Mpc})^3$ for BGS. 
These values approximately match the monopole of the 
power spectrum at $k=0.15h/{\rm Mpc}$ for the respective samples.

The redshifts and all four weights are then randomly sampled from the data catalog and attached to the random catalogs in 
order to match the radial selection function. 
Finally, the catalogs are normalized separately (and all weights are fit for separately) in the 
North and South photometric regions.  

%%%%%%%%%%%%%%%%%%%%%%%%%%%%%%%%%%%%%%%%%%%%%%%%%%
%%%%%%%%%%%%%%%%%%%%%%%%%%%%%%%%%%%%%%%%%%%%%%%%%%

\section{Mock Catalogs}\label{sec:mocks}

%--------------------------------------------------------------- 
%---------------------------------------------------------------  

In our analysis, we utilize  DESI mock galaxy catalogs 
for statistically testing the performance of the BAO fits, as well as
for validating the adopted covariance matrices
in terms of BAO fitting. Here,
we briefly describe the main characteristics
of the various sets of mocks,
along with the 
DESI customization procedure to 
include survey realism.
 
%--------------------------------------------------------------- 
%---------------------------------------------------------------  

\subsection{DESI Mocks: General Description}

We use two different sets of DESI mock galaxy catalogs for the LRG sample: 
one type is directly constructed from $N$-body simulations 
\citep[i.e., \texttt{AbacusSummit};][]{AbacusSummit2021}, 
while a second type is 
based on approximated methods
\citep[i.e., \texttt{EZmocks};][]{EZmocks2021}.

% AbacusSummit Mocks

The $N$-body-based realizations are part of the first official set of DESI mock galaxy catalogs (Alam et al. in preparation) which 
were calibrated based on an early reduction of the One Percent Survey spectroscopic data for LRG 
(see Section \ref{subsec:calibration}).\footnote{The matching data in the final reductions 
are publicly released as part of the  
DESI EDR.}
This set is  made of
$25$ cutsky simulations based on the $2\ihGpc$ \texttt{AbacusSummit} runs.\footnote{\url{https://abacussummit.readthedocs.io/en/latest/}}
The halo occupation distribution (HOD) 
model for LRGs is calibrated using small-scale (below $5 \ihMpc$) wedges in 
combination with large-scale bias evolution where available. 
The LRG mocks are further subsampled to approximately match the $n(z)$ distribution of the specific LRG sample (called `main') 
considered in this paper \citep[see Figure 16 of ][for the `main' selection in the One Percent Survey]{DESIlrgtarget}.
 The mocks implementing the DESI survey geometry and specifics (denoted as `cutsky' mocks)
 are generated using the simulation output near the primary redshift of LRGs that 
we do not report in this paper. The $2 \ihGpc$ box is repeated and then the 
coordinates are converted to sky coordinates.\footnote{The code used to create the cutsky/lightcones can be found at \url{https://github.com/Andrei-EPFL/generate_survey_mocks/}} 

% EZmocks

The approximate mock realizations consist of 1000 \texttt{EZmocks} for LRGs, and 
are built with an elaborated procedure
centered on the Zel’dovich approximation \citep{Zeldovich1970}. They do account for 
stochastic scale-dependent, non-local, and nonlinear biasing contributions:
extensive details on the production methodology
can be found in the original release paper by \cite{Chuang2015}. 
The \texttt{EZmocks} have accurate clustering properties consistent with the
previously described $N$-body-based \texttt{AbacusSummit} realizations -- and nearly indistinguishable
from actual $N$-body solutions -- in terms of one-point, two-point,
and three-point statistics.

% No Mocks for BGS 
 
 We note that we have decided not to use any mocks for the BGS sample,
 primarily for reasons related to a calibration performed with an earlier DESI dataset than
 the one considered in this study.

%--------------------------------------------------------------- 
%---------------------------------------------------------------  

\subsection{DESI Mocks: Masking and Customization} \label{subsec:masks}

%-------------------------------%

\begin{figure}
    \centering
    \includegraphics[width=1.05\columnwidth]{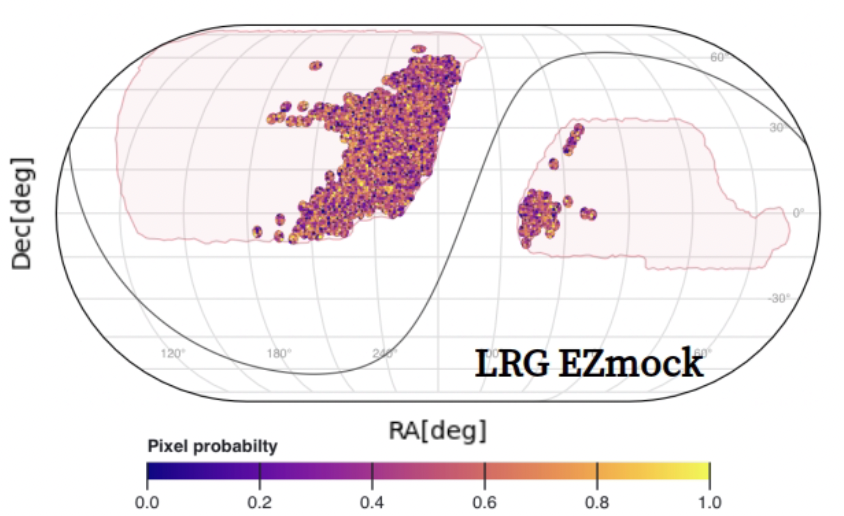}
    \includegraphics[width=1.05\columnwidth]{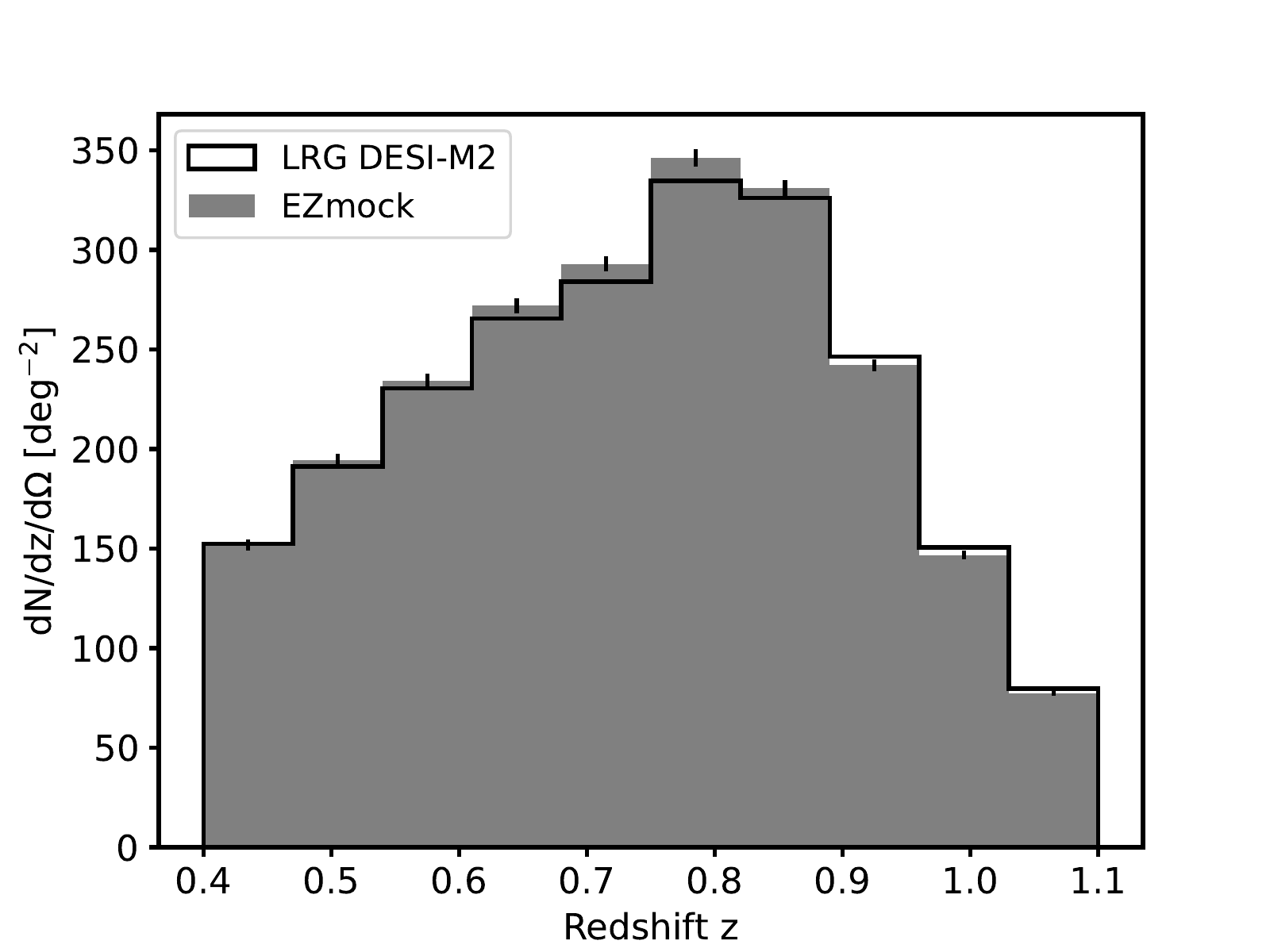}
    \caption{[Top] Map of a single LRG \texttt{EZmock} realization with the pixel probability of the \texttt{HEALPix} mask. 
    [Bottom] Dispersion of $1000$ LRG \texttt{EZmocks}, after application of the different masks described in Section \ref{subsec:masks}, compared with the actual LRG
    \desimtwo\ redshift distribution.}
    \label{fig:mock_prepare}
\end{figure} 

%-------------------------------%

An important step in the mock-making pipeline is the incorporation of
survey realism, namely the characteristics of the 
various synthetic realizations need to accurately
match the properties of the \desimtwo\  sample.
This is achieved via the application of
a succession of masks, as we schematically describe in what follows. 

{\it Survey Masks}. 
We subsample the mocks using a tile mask matching the footprint of \desimtwo\ and match the redshift 
distribution of each target: for LRGs, 
the $n(z)$ distribution is based on \desimtwo\ results (Figure \ref{fig:footprint}), 
while for BGS on the One Percent Survey distribution. This tile mask cuts the data and random to the circular 
region around each tile center.

{\it Intra-tile Geometry}. The geometrical area where DESI targets could have been observed is obtained for the data catalogs following the procedure 
described in Section \ref{sec:LSScat}. In order to analyze all $1000$ mocks, we 
approximate the results of the procedure run on the data using a \texttt{HEALPix} \citep{HEALPix} map built from the  
random catalogs. 
As a reference, we utilize the same random cuts to the target catalog sky area that were the inputs to the LSS catalog process, additionally cutting them 
to be within the tile area previously defined. In every $N_{\rm side} = 1024$ pixel, we count the number of randoms in both the LSS catalog and in the reference randoms. 
The ratio of these counts approximates the small-scale holes in the observed footprint. 
We apply it by sub-sampling the mock data and 
randoms by the fraction in each pixel. In this way, the effective area of the footprint, 
as determined by summing random points assumed to have a constant surface density, 
matches that of the observed data. This is shown as an example in the top panel of Figure \ref{fig:mock_prepare}, 
to be compared with the bottom panel of Figure \ref{fig:footprint}.

{\it Incompleteness Assignment}.  At this stage, we still 
have more simulated galaxies than those observed in the actual DESI data. 
This is because, for most of the locations, only one fiber is available to observe multiple targets. In order to approximate this effect in terms of number 
counts, we simply take the overall assignment completeness of the data in the LSS catalogs, i.e., $N_{\rm observed}/N_{\rm total}$, 
where $N_{\rm total}$ is the number of targets within the \desimtwo\ footprint where observations were possible. This type of completeness 
should vary strongly as a function of the number of overlapping DESI tiles, but we simply apply a constant factor 
(an average of $0.51$ for the $1000$ realizations) given that over $90\%$ of the \desimtwo\ area is covered by only a single tile. 
To this end, the bottom panel of Figure \ref{fig:mock_prepare} shows that the observed \desimtwo\  LRG redshift counts per square degree
match well those obtained from mock data,  after applying the assignment incompleteness factor. 
This procedure is implemented in the  $N$-body based
\texttt{AbacusSummit} realization as well as in the approximate \texttt{EZmocks.} 
Finally, we note that the incompleteness will be modeled more rigorously  
for the forthcoming DESI Y1 analysis.

%--------------------------------------------------------------- 
%---------------------------------------------------------------  

\subsection{DESI Mocks: Calibration} \label{subsec:calibration}

\begin{figure}
    \centering
    \includegraphics[width=1.05\columnwidth]{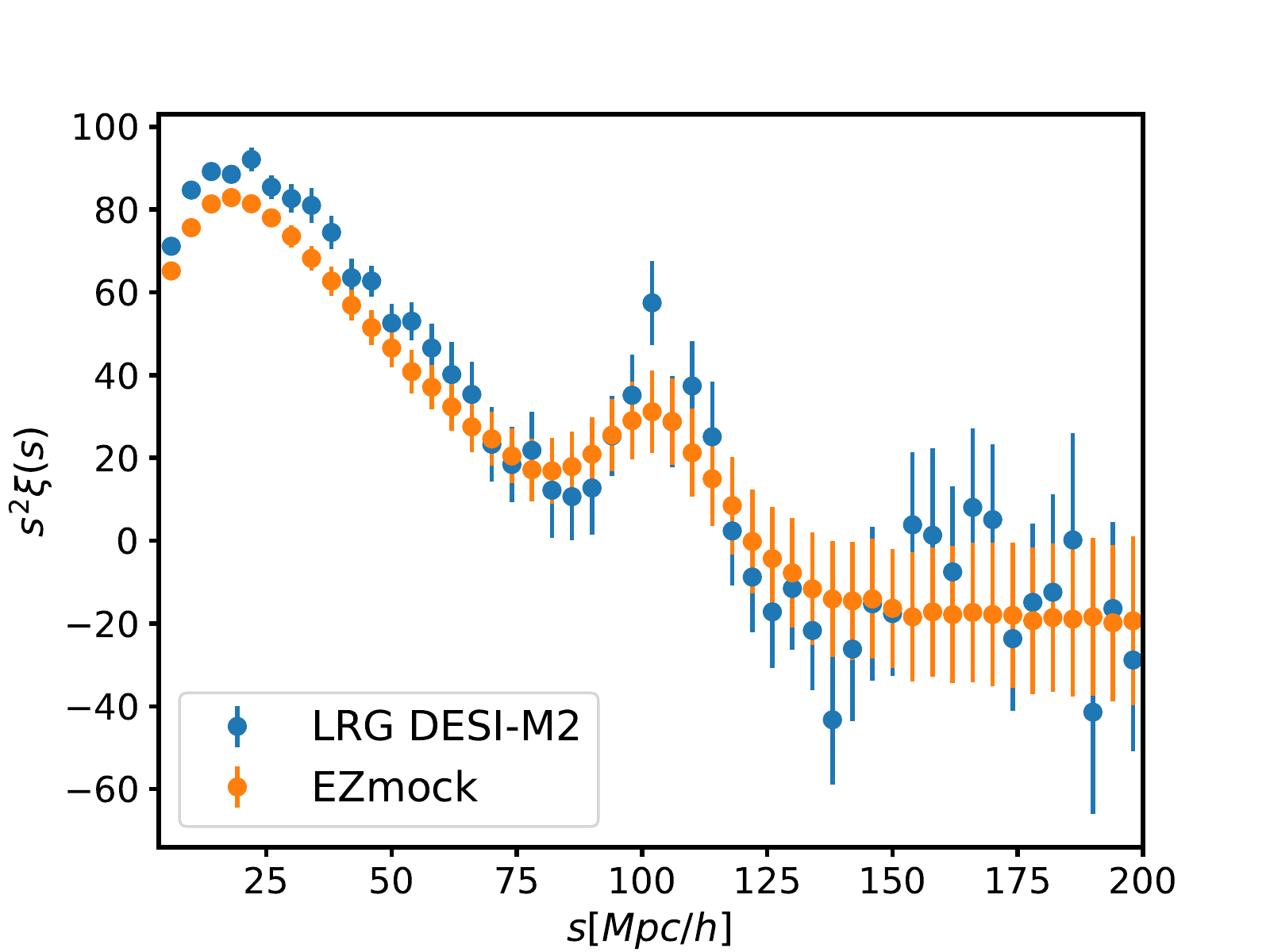} 
    \caption{Monopole of the LRG two-point correlation function before reconstruction, as measured from 
     the \desimtwo\ sample (blue dots) and from the average of 1000 \texttt{EZmocks} (orange dots).
     Data errorbars are obtained from a jackknife covariance directly inferred from \desimtwo\ LRGs, while
     mock errorbars are drawn from the LRG \texttt{EZmock} sample covariance. 
     As mentioned in the main text (Section \ref{subsec:calibration}), the $\sim 10\%$ difference near the $\sim 20~\ihMpc$ peak is not surprising, as
     these mocks were tuned with an earlier version of the DESI data. Hence, in the present work
     mocks are only used for validation purposes.}
    \label{fig:data-mock-LRG-2PCF}
\end{figure}

In terms of clustering properties (see Section \ref{subsec:methods:clustering}), 
the two sets of mocks described here have been 
tuned via survey and completeness masks 
with an earlier version of the DESI LRG clustering 
measurements (i.e., One Percent Survey data)
having a 10\% lower amplitude than the \desimtwo\ LRG sample considered in the current study. 
This can be readily inferred from Figure \ref{fig:data-mock-LRG-2PCF}, where
we contrast the observed LRG clustering in the \desimtwo\ sample 
at $r\sim 20\ihMpc$ with the average clustering of 1000 LRG \texttt{EZmocks}.
For this reason, in the present analysis 
mocks are only used for 
validation purposes, 
and we will be 
adopting semi-analytical semi-empirical covariances rather than mock-based covariances for our primary BAO fits
-- as described in Section \ref{sec:cov}.
In fact, a calibration offset in the two-point clustering (although located outside of the BAO fitting range)
would manifest in a  substantial difference in the covariance between different scales, causing a 17.6\% 
impact on the resulting BAO precision when pairing such a mock-based covariance matrix with the actual data clustering.  

%-------------------------------% 

%-------------------------------% 

%%%%%%%%%%%%%%%%%%%%%%%%%%%%%%%%%%%%%%%%%%%%%%%%%%
%%%%%%%%%%%%%%%%%%%%%%%%%%%%%%%%%%%%%%%%%%%%%%%%%%

\section{Analysis Methods}\label{sec:methods}

%--------------------------------------------------------------- 
%---------------------------------------------------------------  

In this section, we illustrate all of the analysis tools 
adopted in our work, from the 
two-point clustering estimator to the density field reconstruction, 
until the BAO fitting methodology.
In particular, the DESI team is currently studying all aspects of 
the BAO pipeline given the stringent requirements on 
theoretical and observational systematics that will be imposed by a dataset as powerful 
as we expect by the end of the survey.  
For the investigation of this preliminary \desimtwo\ data, and to make contact with 
earlier work on the subject, we choose 
to largely follow the analysis choices made by the BOSS/eBOSS surveys.  
We highlight these choices in what follows, while referring the reader to the original papers for 
more extensive details.

%--------------------------------------------------------------- 
%---------------------------------------------------------------  

\subsection{Two-Point Correlation Function Estimator}  \label{subsec:methods:clustering}

%-------------------------------%

\begin{figure*}
    \centering
    \includegraphics[width=0.99\columnwidth]{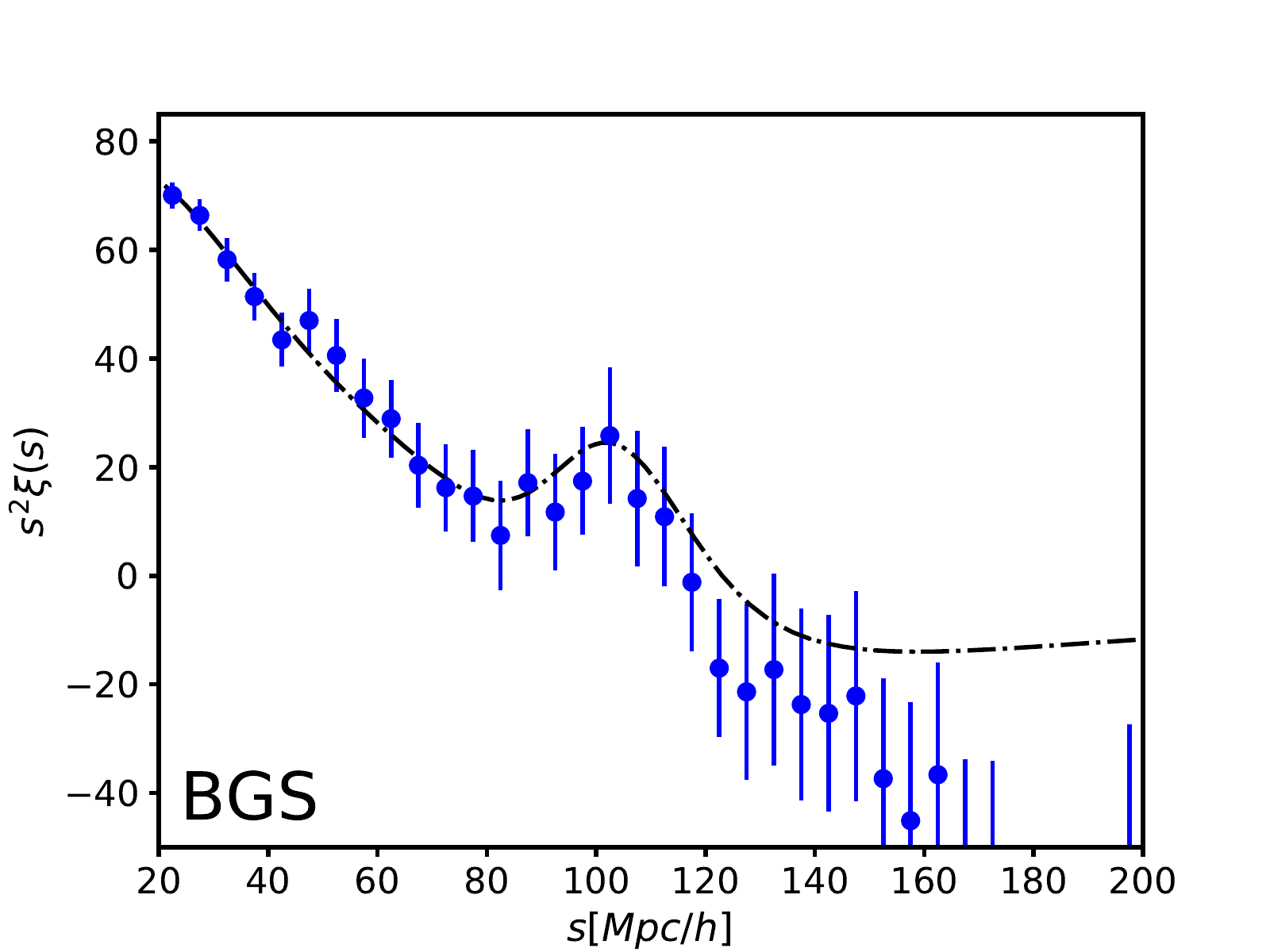} 
    \includegraphics[width=0.99\columnwidth]{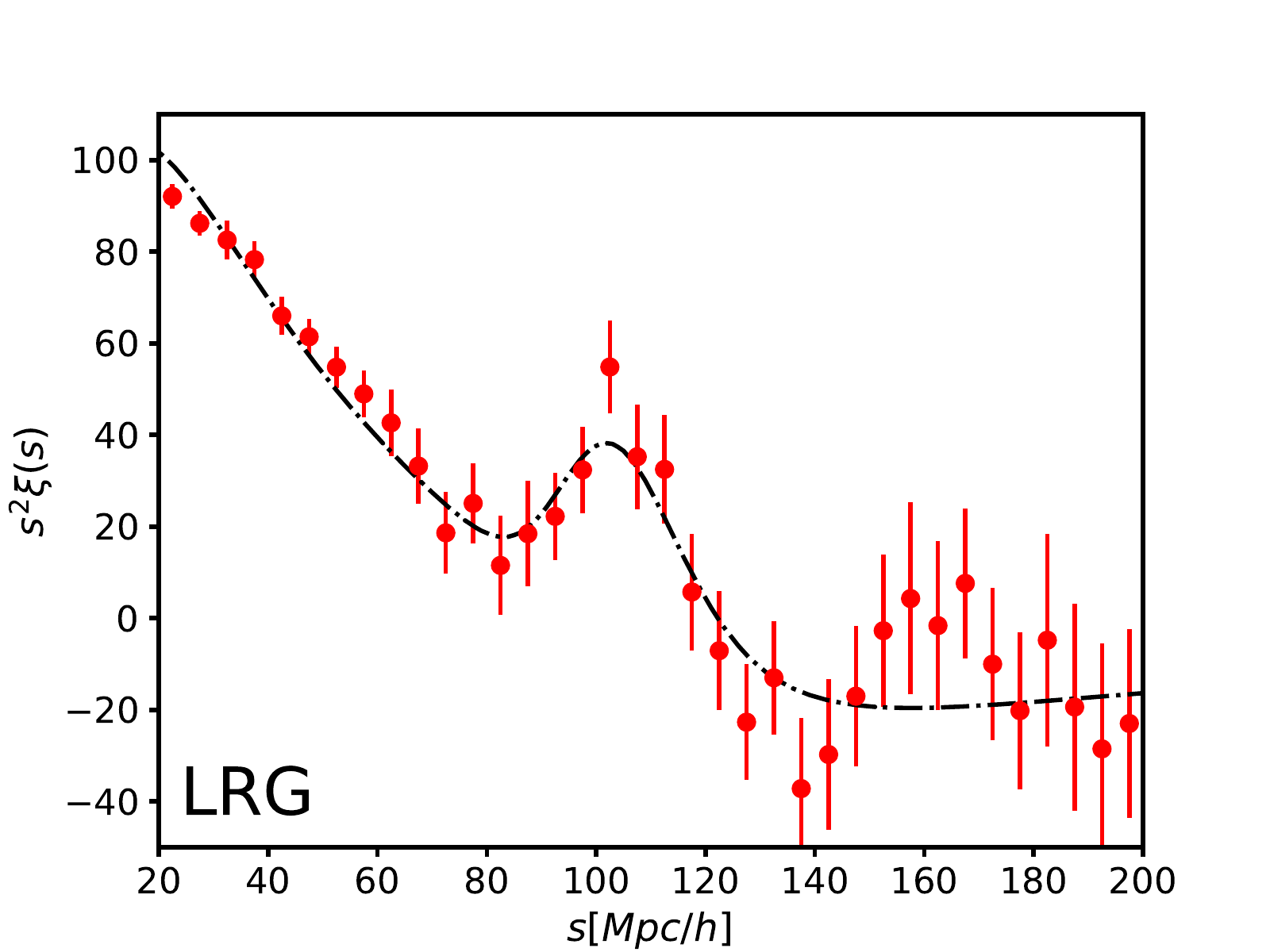} 
    \includegraphics[width=0.99\columnwidth]{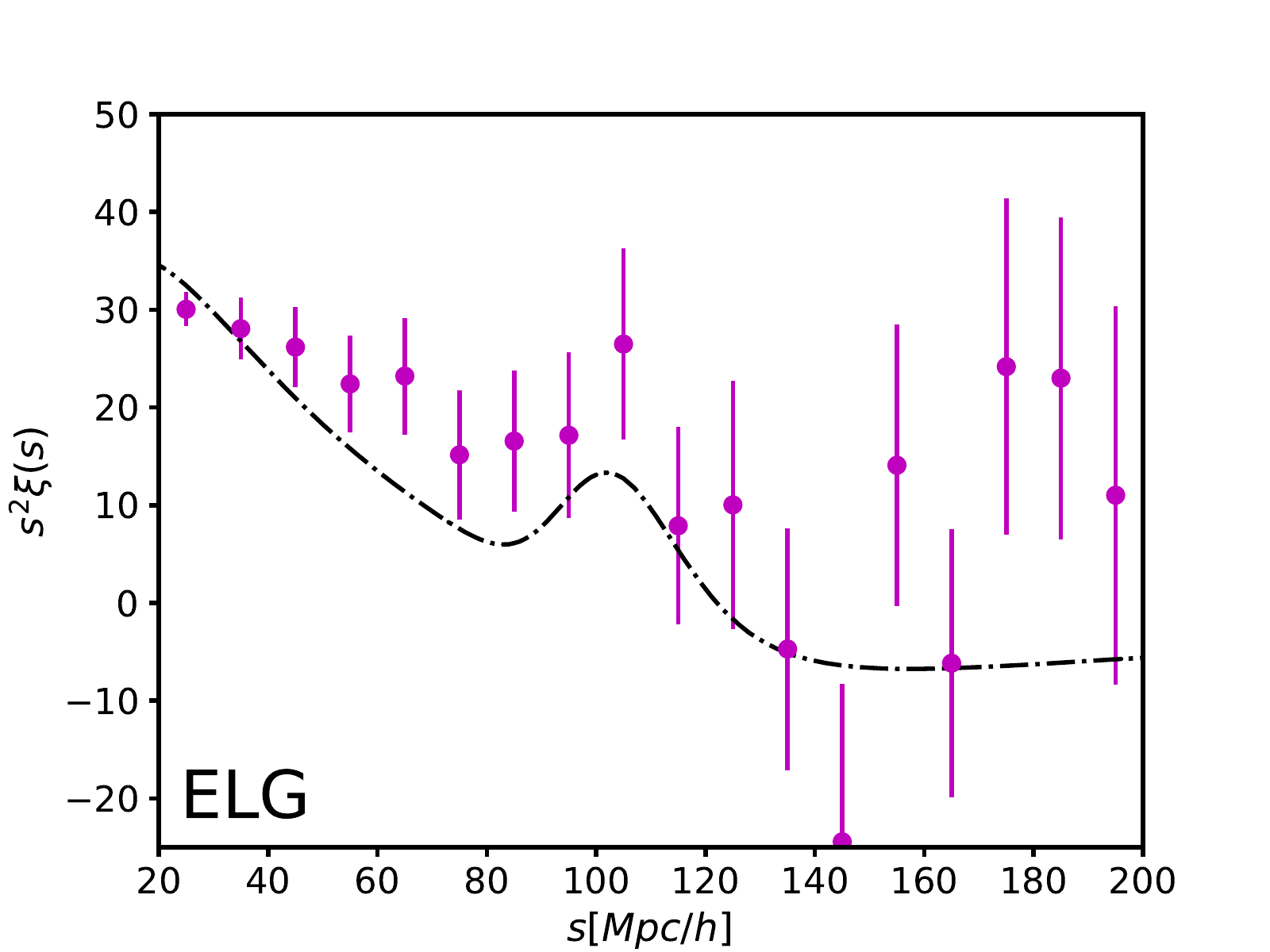}
    \includegraphics[width=0.99\columnwidth]{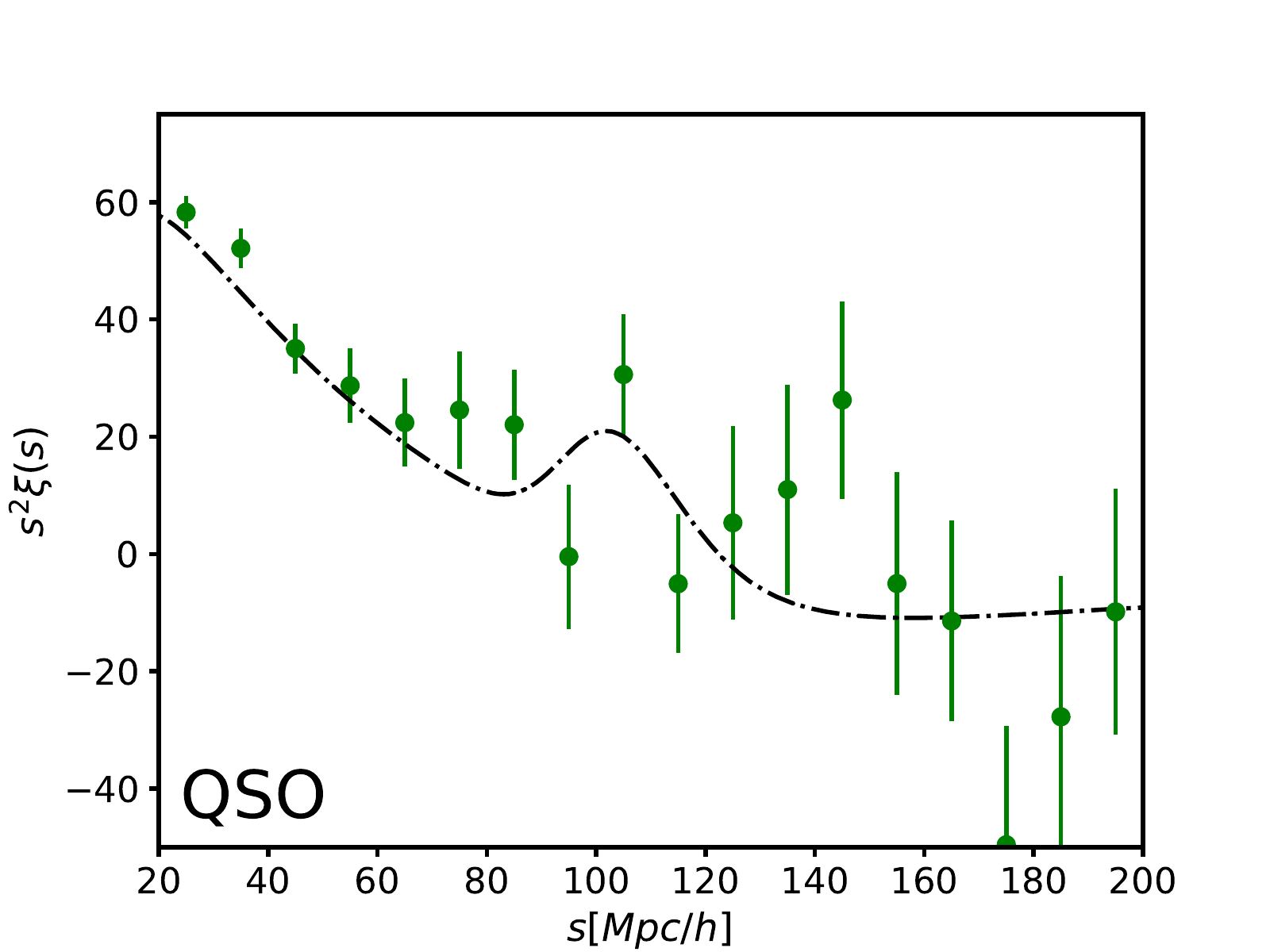}
    \caption{Two-point correlation function measurements of the four DESI tracers, obtained 
    from the \desimtwo\ sample. 
    Errorbars are derived from the diagonal of the corresponding covariance matrixes, although 
    we caution the reader of a significant bin-to-bin correlation in these measurements. 
    Model curves are simple damped linear theory predictions that indicate the expected overall clustering amplitude and BAO damping 
typical at the mean redshift of the target samples.}
    \label{fig:meas2pcf}
\end{figure*}

%-------------------------------%

We compute all of the 
anisotropic redshift-space correlation functions $\hat{\xi}$'s with the well-known 
Landy-Szalay estimator \citep[LS;][]{Landy1993}, namely:
\begin{equation}
\hat{\xi}(s, \mu) = \frac{DD(s, \mu) - 2 DR(s, \mu) + RR(s, \mu)}{RR(s, \mu)},
\label{eq:landyszalay_estimator}
\end{equation}
where $DD(s, \mu)$ and $RR(s, \mu)$ are the normalized weighted number of pairs in the data and random catalogs -- respectively --  
binned as a function of the separation $s$ between two galaxies, $\mu \in [-1, 1]$ is the cosine angle between the 
galaxy pair and the line of sight (LOS), and 
$DR(s, \mu)$ denotes pair counts between data ($D$) and randoms ($R$). 
The LS estimator gets modified when the reconstruction procedure (described in Section \ref{subsec:methods:recon}) is applied. 
In essence, a shifted random catalog (termed $S$) should be used in the numerator of Equation \eqref{eq:landyszalay_estimator} in substitution of $R$,
and one needs to replace $DR$ with $DS$ and $RR$ with $SS$, respectively. We use $200$ $\mu$-bins spanning the interval [$-1,1$] and $4h^{-1}{\rm Mpc}$ $s$-bins for BGS and LRGs.  

The anisotropic correlation function $\hat{\xi}(s, \mu)$  is then integrated over the Legendre polynomials $\mathcal{L}_{\ell}(\mu)$  to obtain the various multipoles; in the current analysis, we only use the monopole, i.e., $\ell=0$:
\begin{align}
\hat{\xi}_{\ell}(s) &= \frac{2 \ell + 1}{2} \int_{-1}^{1} d\mu \hat{\xi}(s, \mu) \mathcal{L}_{\ell}(\mu) \\
&\simeq \frac{2 \ell + 1}{2} \sum_{i} \hat{\xi}(s, \mu_{i}) \int_{\Delta \mu_{i}} d\mu \mathcal{L}_{\ell}(\mu).
\end{align}
In the last equality, we have made explicit the discrete sum over $\mu$-bins, weighted by the analytic integral of $\mathcal{L}_{\ell}(\mu)$ 
over each $\mu$-bin having width $\Delta \mu_{i}$.\footnote{Such a summation scheme, contrary to weighting $\hat{\xi}(s, \mu_{i})$ by  $\mathcal{L}_{\ell}(\mu_{i}) \Delta \mu_{i}$, 
ensures that the $\ell > 0$ multipoles are exactly zero if $\hat{\xi}(s, \mu)$ remains constant as a function of $\mu$.}

All of the two-point correlation function 
calculations are performed with the Python package \texttt{pycorr},\footnote{\url{https://github.com/cosmodesi/pycorr}} 
which wraps a modified version\footnote{\url{https://github.com/adematti/Corrfunc}} of the \texttt{Corrfunc} package 
\citep{Corrfunc2,Corrfunc1}.\footnote{\url{https://github.com/manodeep/Corrfunc}}

Figure \ref{fig:meas2pcf} shows the observed two-point correlation functions of the four tracers discussed in Section \ref{subsec:targets},
contrasted with simple damped linear theory models that indicate the expected overall clustering amplitude and BAO damping 
typical at the mean redshift of the target samples. 
From LRGs and BGS, we observe a local bump near the expected location of the 
BAO peak. Moreover, while BGS observations appear to lie systematically below the model curve at scales greater than $120 h^{-1}{\rm Mpc}$,
this is simply because there are fewer modes  at larger separations in these early DESI data. Therefore, 
they are highly correlated and thus the amplitude of the two-point correlation function decreases at those scales.\footnote{
In addition, note that we  only fit up to $150h^{-1}{\rm Mpc}$  for the BAO analysis, where the corresponding linear theory prediction 
is still consistent with observations -- within errorbars.}
For ELGs and QSOs, the amplitude of the observed clustering appears consistent with the theoretical expectations (within errors),
although it is challenging to identify a clear BAO-like signature. Indeed, we do not expect a BAO detection from ELGs and QSOs of the \desimtwo\ sample, 
given the small survey volume in combination with the low completeness (ELGs) and high shot noise (QSOs). 

%--------------------------------------------------------------- 
%---------------------------------------------------------------  

\subsection{Density Field Reconstruction} \label{subsec:methods:recon}

We apply the density field reconstruction technique 
\citep{Eisenstein2007:astro-ph/0604362v1} on the observed galaxy density fields in order to partially recover the 
BAO feature that has been degraded due to structure growth and redshift space distortions (RSD).
To do so, we follow the iterative procedure described in \citet{Burden2015:1504.02591v2}, as implemented in the \texttt{IterativeFFTReconstruction} 
algorithm of the \texttt{pyrecon} package\footnote{\url{https://github.com/cosmodesi/pyrecon}} with the \texttt{RecIso} convention.\footnote{\texttt{RecIso} is a choice to remove 
the large-scale anisotropy due to redshift-space distortions in the process of reconstruction \citep{PadRecIso,SeoBAOmodel2016}.} 
The density contrast field is smoothed by a Gaussian kernel of width $15 \ihMpc$ and three iterations are performed, 
assuming an approximate growth rate and the expected bias for each sample.
The choice of these reconstruction conditions along with the assumed fiducial 
cosmology were shown to have a very marginal impact on BAO measurements in earlier galaxy survey samples -- i.e., 
see \cite{VargasMagana2016:1610.03506v2} and \cite{Carter2020:1906.03035v1}.

%--------------------------------------------------------------- 
%---------------------------------------------------------------  

\subsection{BAO Fitting Methodology} \label{subsec:methods:fitting}

%-------------------------------%

\begin{table*}
\centering
\caption{Covariance matrices utilized in this work.}
\begin{tabular}{c|c|c}
\hline\hline
{\bf Name} & {\bf Tracer} & {\bf Notes}  \\ \hline
\desimtwo-EZ & LRG & Constructed from \texttt{EZmock} clustering \\
\rascalc-EZ & LRG & \rascalc\ calibrated on \texttt{EZmock} clustering \\
\rascalc-LRG & LRG & \rascalc\ calibrated on \desimtwo\ LRG clustering \\
\rascalc-BGS & BGS & \rascalc\ calibrated on \desimtwo\ BGS clustering \\
\hline \hline
\end{tabular}
\label{tab:covariance}
\end{table*}

%-------------------------------%

We employ the same BAO fitting pipeline that has been previously applied to a large number of  BOSS and eBOSS analyses
\citep{Ross2017,Ata2018,HouDR16QSO,RaichoorDR16ELG}.\footnote{\url{https://github.com/ashleyjross/BAOfit}}
The accuracy of such methodology was demonstrated to be sufficient at the precision demanded by BOSS/eBOSS data, 
especially for determining the isotropic BAO scale. 
However, advances are expected to be necessary in order to meet the exquisite precision expected 
for DESI Y5, and thus an improved DESI pipeline 
is currently under development and will be presented along with the DESI Y1 analyses.

In this study, we only fit for the 
monopole of the correlation function, hence for the isotropic scaling parameter $\alpha$. 
Our BAO pipeline is `template-based', and it essentially coincides with the algorithm introduced by \cite{XuBAOfit}. 
However, the BAO templates are generated via the formulae defined by 
Equations 9-13 of \cite{Ross2017}, where the linear power spectrum is split 
into a BAO and a no-BAO components, and damping is added solely to the BAO that depends on the LOS angle. 

The templates require a choice of four parameters that are kept fixed during the fitting process, 
namely: $\beta$, $\Sigma_{\rm s}$, $\Sigma_{||}$, $\Sigma_{\perp}$. 
These determine, respectively, the degree of anisotropy with respect to the LOS in linear RSD \citep{Kaiser1987}, 
the degree of streaming velocity, the degree of radial BAO damping, 
and the degree of transverse BAO damping. 
Such parameters are fixed separately for different samples and the pre- or post-reconstruction fits. 
For galaxy velocities, we set $\beta = 0.4$, $\Sigma_{\rm s} = 3\ihMpc$ for both samples;
such choices guarantee an approximate match to the anisotropic clustering as measured from the 
\desimtwo\ data, although their impact in the fitting process is essentially negligible since we only 
fit for the monopole.
The BAO damping parameters for post-reconstruction are fixed to $\Sigma_{||}$, $\Sigma_{\perp}$ = 3, 5 $\ihMpc$ for both samples, roughly consistent with those used/determined in previous studies (e.g., \citealt{SeoBAOmodel2016,Ross2017,VargasMagana2016:1610.03506v2,Bautista2021}). For pre-reconstruction, we set these to 6, 10 $\ihMpc$ 
for the BGS sample (again, roughly consistent with the pre-reconstruction results from \citealt{SeoBAOmodel2016}) and reduce them to 4, 8 $\ihMpc$ for pre-reconstruction LRGs.
This evolution in the pre-reconstruction values roughly corresponds to the change in the linear growth factor between the effective redshifts $z_{\rm eff}$ of the two samples,  
noting that the BAO damping is expected to scale with the amount of non-linear structure growth, which approximately 
scales with the linear  growth factor \citep{SeoBAOmodel2016}. 
The post-reconstruction values are smaller and constant as reconstruction helps to reduce the effect of non-linearities (hence, smaller damping values), and the degree of remaining non-linearity does not depend strongly on the initial degree (hence the independence of redshift).
The impact of fixing these choices was already shown to be negligible at the precision of BOSS DR12 \citep{Ross2017},
and therefore is also not a concern in the present work.

The procedure just described produces a theory template, $\xi_0$. 
Subsequently, the data is fit against this template evaluated with a scaling parameter $\alpha$, a free amplitude, 
and a polynomial with three nuisance terms:
\begin{equation}
    \xi_{\rm mod}(s) = B\xi_{0,\rm t}(s\alpha) + A_0 + A_1/s + A_2/s^2.
\end{equation}
The polynomial has been shown to account for any difference between the broad-band shape of the template $\xi_{0, \rm t}$ and the measured $\hat{\xi}_0$; 
e.g., due either to cosmology or to observational systematics. 
The model is evaluated at the $s$ of the data bin assuming a spherically symmetric distribution.\footnote{For our binsize of 4$\ihMpc$, this corresponds to a 0.03\% effect compared to just using the bin center.} 
The $\chi^2(\alpha)$ is computed on a grid of spacing $0.001$ in $\alpha$, where the minimum $\chi^2$ at each grid point is determined 
by varying $B,A_0,A_1,A_2$. We note that 
the various $\chi^2$ are inferred 
from the data vector $\vec{D}$ and covariance matrix $\boldsymbol{C}$ via $\chi^2 = \vec{D}\boldsymbol{C}^{-1}\vec{D}^{t}$, as routinely done. 
Our data vectors are always selected to have $50<s<150\ihMpc$. 

Finally, the derived likelihood on the value of $\alpha$ can be used to constrain cosmological models via
\begin{equation}
    \alpha = \frac{D_{\rm V}(z)r^{\rm fid}_{\rm d}}{D^{\rm fid}_{\rm V}(z)r_{\rm d}}
\end{equation}
and 
\begin{equation}
    D_{\rm V}(z) = \left[cz(1+z)^2H(z)^{-1}D^2_{\rm A}(z)\right]^{1/3},
\end{equation}
where $H(z)$ and $D_{\rm A}(z)$ are evaluated at an effective redshift 
of the data sample being tested.

In closing this part, we highlight that 
while the methodology adopted here is 
largely equivalent to the one exploited in previous BOSS/eBOSS analyses, 
the version of the BAO pipeline used in this work has been fully updated to be compatible with DESI code packages 
assuming generic cosmological backgrounds and primordial/linear power spectrum calculations:
such effort is carried out within the \texttt{cosmodesi} framework.\footnote{\url{https://github.com/cosmodesi/BAOfit\_xs/}} 
To this end, 
the most significant change specific to the BAO fitting procedure is how we isolate the BAO feature, namely by splitting the input linear power spectrum 
into a smooth function with no-BAO and another one that is pure BAO. 
To achieve such splitting, we apply the technique described in \cite{Wallisch2018} and coded in the \texttt{bao\_filter} module\footnote{\url{https://github.com/cosmodesi/cosmoprimo/blob/main/cosmoprimo/bao\_filter.py}} of the \texttt{cosmoprimo} package.
The impact of this change in filtering the BAO feature is less than $\sim 0.1\%$ on the measured value of $\alpha$.  

%%%%%%%%%%%%%%%%%%%%%%%%%%%%%%%%%%%%%%%%%%%%%%%%%%
%%%%%%%%%%%%%%%%%%%%%%%%%%%%%%%%%%%%%%%%%%%%%%%%%%

\section{Covariance Matrices}\label{sec:cov}

%--------------------------------------------------------------- 
%---------------------------------------------------------------  

In this section, we briefly address covariance matrices, 
and in particular the construction, calibration, and validation
of semi-analytical semi-empirical covariances on mock data -- eventually adopted for the BAO fitting procedure.

%--------------------------------------------------------------- 
%---------------------------------------------------------------  

\subsection{Covariance Matrices: Types and Conventions}\label{subsec:cov-types-conventions}  

%-------------------------------% 

\begin{figure*}
\centering
\includegraphics[width=0.3\textwidth]{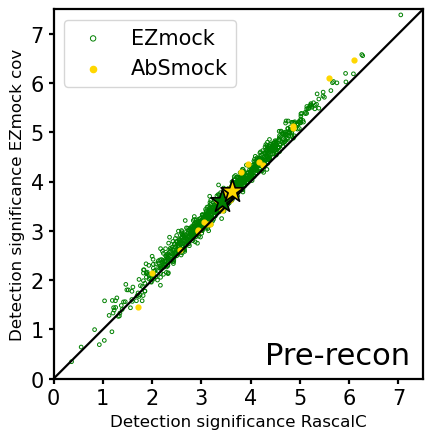}
\includegraphics[width=0.32\textwidth]{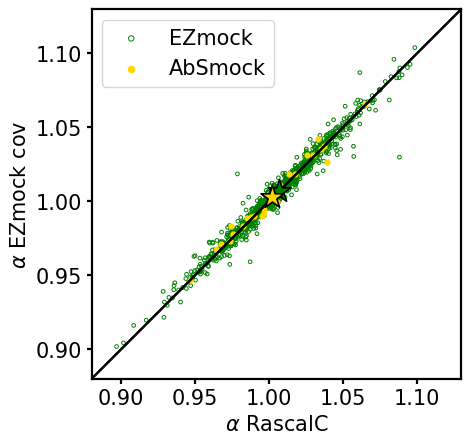}
\includegraphics[width=0.32\textwidth]{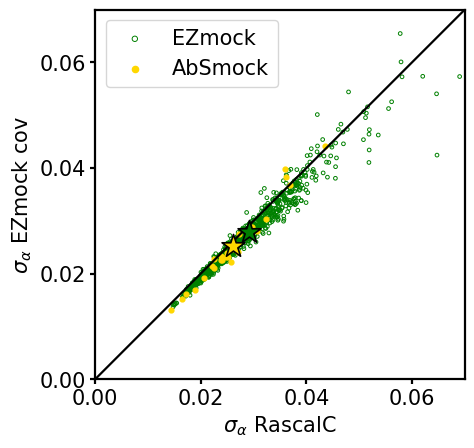}
\includegraphics[width=0.3\textwidth]{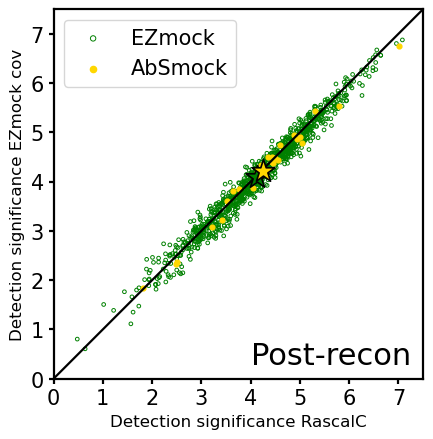}
\includegraphics[width=0.32\textwidth]{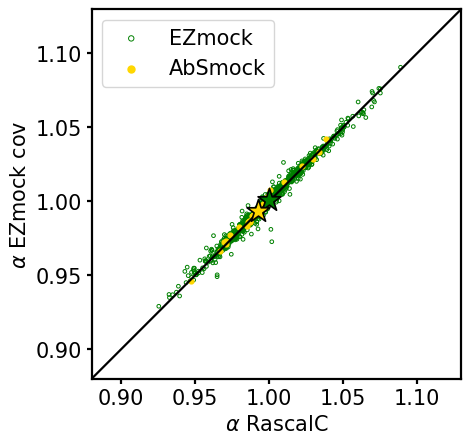}
\includegraphics[width=0.32\textwidth]{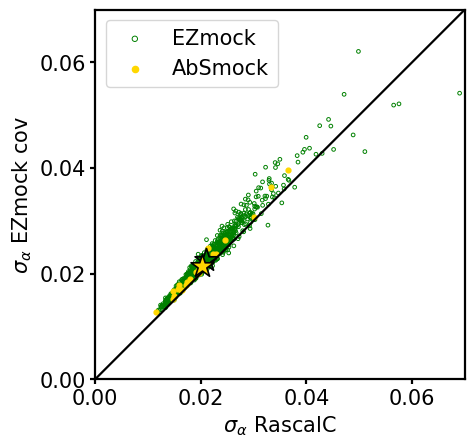}
\caption{Scatter plots showing the BAO detection significance in units of standard deviations (left panels), the $\alpha$'s (middle panels), 
and the $\sigma_{\alpha}$ values (right panels) related to the 
validation procedure of the \texttt{RascalC}-based covariance. 
BAO best fits are performed on a mock-by-mock basis, using 1000 LRG \texttt{EZmocks} (open green dots) 
and 25 \texttt{AbacusSummit} LRG realizations (filled yellow points). Stars of the same color are 
averages over the corresponding entire set of realizations. Top panels refer to pre-reconstruction measurements, 
while bottom panels display post-reconstruction quantities. Individual BAO fits using the \texttt{RascalC} 
covariance calibrated on \texttt{EZmocks} (i.e., \texttt{RascalC}-EZ; $x$-axes) 
are contrasted with those performed adopting the \texttt{EZmock} covariance (i.e., \desimtwo-EZ; $y$-axes). 
As evident from the figure, the narrow scatter along the diagonal 
implies that both covariances produce compatible results, validating the usage of 
\texttt{RascalC}-based covariances for our primary  \desimtwo\ BAO fits.}
\label{fig:scatter_Rascal_EZ_LRG}
\end{figure*}

%-------------------------------% 

The primary BAO fits performed in our main analysis are 
obtained with semi-analytical semi-empirical covariance matrices, generated by the {\tt RascalC} code \citep{RascalC}.
As mentioned in Section \ref{subsec:calibration}, this choice is mainly
driven by the fact that the galaxy mocks available at the time of this study --
implementing all of the \desimtwo\ survey characteristics --
have been calibrated with an earlier version of the DESI LRG clustering (i.e., the One Percent Survey), 
rather than with the
LRG clustering as measured directly from the \desimtwo\ dataset considered here.
Nevertheless, we also construct a numerical covariance 
from 1000 LRG \texttt{EZmocks} (termed `\desimtwo-EZ'), and use it for validating  the LRG 
{\tt RascalC} semi-analytical covariance 
in terms of BAO fitting.  
In order to calibrate the {\tt RascalC} covariance matrix for LRGs, 
besides the previous  
numerical covariance, we also 
utilize a covariance directly constructed from the jackknife estimates of the LRG sample under consideration.  
Once the {\tt RascalC} semi-analytical covariance is validated (in terms of BAO fits) for the LRG sample,
we build a similar semi-empirical covariance for BGS galaxies  and 
calibrate it using jackknife estimates obtained from the corresponding BGS dataset. 
Table \ref{tab:covariance} reports all of the covariance matrices utilized in this work.
Specifically, in terms of name conventions, we
indicate with  {\tt `RascalC}-EZ' the semi-analytical covariance
based on the \texttt{EZmocks} LRG clustering,
with {\tt `RascalC}-LRG' the semi-empirical covariance
calibrated on the \desimtwo\ LRG clustering measurements (including jackknife), and with
{\tt `RascalC}-BGS' the one based on
the \desimtwo\ BGS clustering. 

Next, we provide some additional 
information on {\tt RascalC}-based covariances,
and then present the BAO fitting validation tests performed on  
the mock LRG sample.  

%--------------------------------------------------------------- 
%---------------------------------------------------------------  

\subsection{{\tt RascalC} Covariances} \label{subsec:RascalCcov}  
 
%-------------------------------% 

The semi-analytical semi-empirical covariance matrices
used in our analysis are obtained via the
publicly available code \rascalc\ \citep{RascalC,RascalC-DA02}.\footnote{\url{https://github.com/oliverphilcox/RascalC}}  
The procedure to construct such covariances
only requires a two-point correlation function as input (along with its optional jackknife estimates), 
and a random catalog.
The \rascalc\ algorithm integrates the Gaussian terms for the covariance matrix  
using importance sampling from the set of random points.
It then progressively changes the amount of shot noise, which has the effect of empirically rescaling those terms.
The optimization of the shot noise level is performed on separate jackknife covariance estimates.
Once the optimal shot noise level is determined (i.e., its best-fit value), 
a rescaling based on such best-fit is applied
to finally obtain the full covariance matrix terms.

For the construction of the  {\tt RascalC}-LRG and  
{\tt RascalC}-BGS covariances (i.e., the covariances calibrated on the \desimtwo\ dataset), 
the input correlation functions are measured directly from the \desimtwo\ 
LRG and BGS samples, respectively. In addition, 
60 jackknife regions are assigned based on data points with a K-means subsampler.
For the pre-reconstruction case, 
the \rascalc\  code is run on 10 random catalogs separately, and the integration results are finally averaged.
Building the post-reconstruction covariances calibrated on the \desimtwo\ dataset requires some
additional steps, described in detail in \cite{RascalC-DA02}. In essence,
such procedure depends upon the usage of 
non-shifted random catalogs for normalization, shifted random catalogs for sampling, and 
a slightly different two-point correlation function than the
familiar LS estimator (i.e., Equation \ref{eq:landyszalay_estimator}). 
In this latter case,  the code is run on 20 random catalogs separately, and 
integration results are eventually averaged.

In order to build the {\tt RascalC}-EZ covariance, we use
instead the averaged pre- and post-reconstructed LRG \texttt{EZmock} 
correlation functions obtained from 1000 \texttt{EZmocks}, 
without jackknife estimates, and with no shot-noise rescaling (Gaussian). 
We run the {\tt RascalC} code 
on 10 concatenated random catalogs in pre-reconstrution, and on
20 randoms for the post-reconstruction case. 

As concluding remarks for this section, we note that, by construction,
the LRG {\tt RascalC} data covariance (i.e., \texttt{RascalC}-LRG) gives larger error bars than the 
\desimtwo-EZ sample covariance, while the
{\tt RascalC} covariance based on the \texttt{EZmock} clustering  
(i.e., \texttt{RascalC}-EZ)  is consistent with the \desimtwo-EZ 
sample covariance. A more detailed assessment on the performance of 
 {\tt RascalC}-based covariances is 
presented in \cite{RascalC-DA02}.

%--------------------------------------------------------------- 
%--------------------------------------------------------------- 

\subsection{Validation of {\tt RascalC} Covariances for BAO Fitting}

%-------------------------------% 

Before performing BAO fits on the \desimtwo\ dataset,  
we validate our LRG {\tt RascalC} covariance
against a set of approximate LRG \texttt{EZmocks} and $N$-body based \texttt{AbacusSummit} realizations. 
While the following tests are performed on LRGs, 
we note that the validation procedure is general and would apply to any tracers.

Specifically,  we first compute the two-point clustering statistics of 1000 LRG \texttt{EZmocks}
and 25 \texttt{AbacusSummit} LRG realizations with the estimator presented
in Section \ref{subsec:methods:clustering}, adopting default FKP weights. 
We then apply the reconstruction algorithm detailed in Section \ref{subsec:methods:recon}
to all of the mocks, assuming a smoothing scale of $15h^{-1}{\rm Mpc}$. 
After, we build the pre- and post-reconstruction \texttt{EZmock} covariance (i.e., \desimtwo-EZ; Table \ref{tab:covariance}), 
and a {\tt RascalC} covariance (i.e., RascalC-EZ; Table \ref{tab:covariance}), which is 
based on the exact average clustering inferred from the entire set of pre- and post-reconstruction LRG 
\texttt{EZmocks}. Finally, we use 
both covariances to fit 1000 \textit{individual} \texttt{EZmocks} as well as 
25 \textit{individual} \texttt{AbacusSummit} realizations
with the fitting procedure explained in Section \ref{subsec:methods:fitting},
within the spatial range $50-150h^{-1}{\rm Mpc}$.\footnote{Note that the Percival factor \citep{Percival2014} 
has been applied to the \texttt{EZmocks}.} 
In essence, we quantify the 
BAO best fits and relative errors on a mock-by-mock basis. 

Figure \ref{fig:scatter_Rascal_EZ_LRG} shows the results of such a validation test.
Top panels refer to pre-reconstruction measurements, while bottom panels display
post-reconstruction quantities. 
From left to right, we report  the 
BAO detection significances in units of standard deviations, and the $\alpha$ and $\sigma_{\alpha}$ values
for all of the individual fits performed to the two sets of mocks
using the \texttt{RascalC}-EZ covariance ($x$-axes), against the corresponding values obtained 
with the \desimtwo-EZ covariance ($y$-axes). 
Open green dots display LRG \texttt{EZmocks} measurements, while filled yellow
dots are for \texttt{AbacusSummit} LRG synthetic catalogs. 
Stars of the same color represent averages over the corresponding 
entire set of realizations.
As evident from the figure, the scatter along the  
diagonal  is quite narrow (both for the pre- and post-reconstruction cases), 
implying that the two covariances
produce compatible results. 
Moreover, the average values in the various panels
are almost overlapping, strongly 
confirming the consistency between {\tt RascalC} and mock sample covariances. 
 
\begin{figure}
\centering
\includegraphics[width=0.44\textwidth]{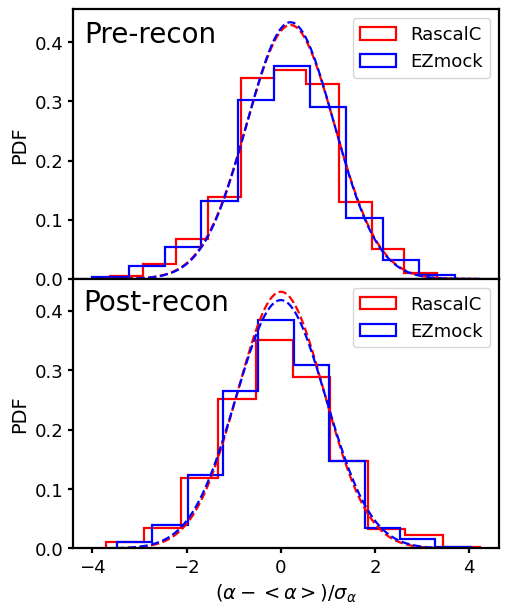}
\caption{Histograms of ($\alpha - \langle \alpha \rangle)/ \sigma_{\alpha}$, with $\langle \alpha \rangle$ the mean of the scaling parameter, measured from the $\xi(s)$'s of the pre- (top panel) and
post-reconstruction (bottom panel) LRG mocks. Measurements are performed on 1000 LRG \texttt{EZmocks}, assuming the \texttt{RascalC}-EZ covariance (red lines and histograms), or a 
numerically-based \desimtwo-EZ covariance (blue lines and histograms). Results are then compared with Gaussian distributions, showing good agreement, and indicating that the values of $\sigma_\alpha$ we measure from the $\chi^2$ distribution are faithful descriptors 
of the error on $\alpha$ measured by fitting $\xi(s)$. This test represents a further validation
of our semi-analytical \texttt{RascalC}-EZ covariance, which produces
results compatible with the numerical one.}
\label{fig:K-S_LRG}
\end{figure}

A further validation test is reported in Figure \ref{fig:K-S_LRG}, where
we show the histograms of ($\alpha - \langle \alpha \rangle)/ \sigma_{\alpha}$, with $\langle \alpha \rangle$ the mean of the scaling parameter, 
measured from the $\xi(s)$'s of the pre- (top panel) and
post-reconstruction (bottom panel) LRG mocks. This quantity represents an approximation for the
signal-to-noise ratio (SNR) of the BAO measurement. 
Here, we use the $\chi^2$ test to assess the validity of the covariances.
In essence, we compare the observed scatter in the best-fitting $\alpha$ for the 1000 LRG \texttt{EZmocks}
to the $\sigma_{\alpha}$ estimated in each individual fit from the $\Delta \chi^2 (\alpha)$ curve. 
Red lines and histograms refer to measurements performed on the
\texttt{EZmock} set using the \texttt{RascalC}-EZ covariance,
while blue lines and histograms correspond to analogous measurements 
done assuming the numerically-based \desimtwo-EZ covariance. 
Results are compared with Gaussian distributions,
showing good agreement, as
confirmed by near-zero Kolmogorov-Smirnov (K-S) 
$D_n$ tests. Moreover, 
the corresponding $p$-values imply that  
our values are drawn from a Gaussian distribution, 
and that the values of $\sigma_\alpha$ we measure from the 
$\chi^2$ distribution are faithful descriptors 
of the error on $\alpha$ measured by fitting $\xi(s)$. 
Once again, this test represents another confirmation of the validity 
of our semi-analytical semi-empirical \texttt{RascalC}-EZ covariance, which produces
results compatible with the numerical case.

%-------------------------------% 

%-------------------------------% 

\begin{figure*}
\centering
\includegraphics[width=0.3\textwidth]{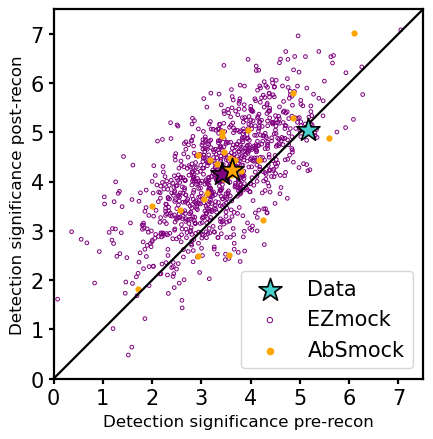}
\includegraphics[width=0.325\textwidth]{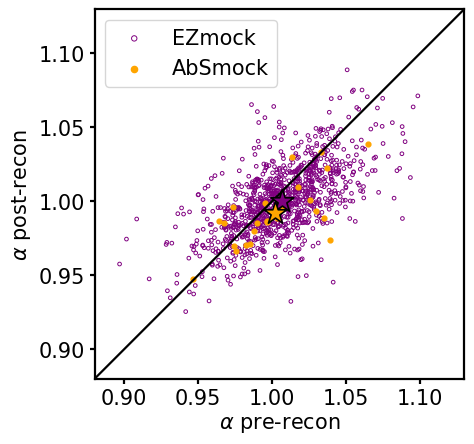}
\includegraphics[width=0.325\textwidth]{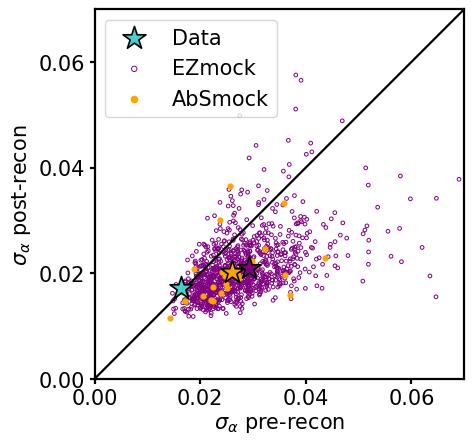}
\caption{Scatter plots for the BAO detection significance in units of standard deviations (left panel), $\alpha$'s (middle panel), 
and $\sigma_{\alpha}$'s (right panel), addressing 
the effect of reconstruction on the BAO fitting procedure for the LRG sample. Results are 
obtained from  individual correlation function fits performed to 1000 LRG \texttt{EZmocks} 
(open purple dots) and 25 \texttt{AbacusSummit} LRG synthetic catalogs (filled orange dots). 
Stars of identical color represent averages over the corresponding entire set of realizations. 
Pre-  ($x$-axes) and post-reconstruction ($y$-axes) measurements are compared,
adopting the \texttt{RascalC-EZ} covariance. While 
reconstruction increases considerably the BAO detection significance for the mocks, the same 
procedure applied to the \desimtwo\ LRG sample produces only marginal effects: the cyan
stars in the left and right panels are   \desimtwo\ LRG measurements obtained 
via a \texttt{RascalC} covariance calibrated directly on LRGs, pointing to a `lucky' realization in 
the upper right corner of the significance scatter plot (left panel), 
and in the lower left corner of the $\sigma_{\alpha}$ plot (right panel), respectively. 
This situation is similar to those reported for BOSS CMASS LRGs \citep{BOSSDR9BAO}, 
and for the eBOSS LRG sample \citep{Bautista2021}.}
\label{fig:scatter_pre-post_LRG}
\end{figure*}

%-------------------------------% 

In summary, the tests performed in this section
clearly prove that using a {\tt RascalC}-based covariance returns unbiased and consistent estimates 
when compared to results obtained with the numerical 
\texttt{EZmock} covariance. We can then safely proceed to tune our \texttt{RascalC} covariance to match the clustering
inferred from the \desimtwo\  datasets, 
and perform the key BAO fits on the LRG and BGS samples -- as we describe in the next section.

%%%%%%%%%%%%%%%%%%%%%%%%%%%%%%%%%%%%%%%%%%%%%%%%%%
%%%%%%%%%%%%%%%%%%%%%%%%%%%%%%%%%%%%%%%%%%%%%%%%%%

\section{Key Results: BAO Signal Detection}\label{sec:results}

%--------------------------------------------------------------- 
%--------------------------------------------------------------- 
 
In this section, we present the main results of our \desimtwo\ analysis, 
and assess the precision and detection statistics of the BAO feature both in the LRG and BGS samples. 
We do not report here the best fit BAO scales as inferred from actual data,
since the cosmology is intentionally kept blinded. Our focus is primarily on LRGs, as they are characterized by 
the highest SNR among the four \desimtwo\ tracers.    

%--------------------------------------------------------------- 
%--------------------------------------------------------------- 

\subsection{BAO Reconstruction Efficiency}

%-------------------------------% 

Before performing BAO fits to the \desimtwo\ dataset, we
first address the effect of BAO reconstruction on the
BAO fitting procedure --  focusing on the LRG sample.
In Figure \ref{fig:scatter_pre-post_LRG}, we report
the  BAO detection significances expressed in units of standard deviations, as well as the $\alpha$ and $\sigma_{\alpha}$ values 
from all of the individual correlation function fits performed to the two sets of LRG mocks previously considered. 
Specifically, we compare pre-  ($x$-axes) and post-reconstruction ($y$-axes) measurements, 
obtained by adopting the \texttt{RascalC}-EZ covariance. 
Open purple dots display \texttt{EZmocks} LRG results, while filled orange
dots are for the \texttt{AbacusSummit} LRG synthetic catalogs. 
Similarly to Figure \ref{fig:scatter_Rascal_EZ_LRG}, stars of identical color 
represent averages over the corresponding entire set of realizations.
As evident from the scatter plot, the BAO detection significance (reported in the left panel)
increases considerably after reconstruction,  and
the $\alpha$'s of the mocks are closer to unity after reconstruction (central panel), 
as expected.\footnote{See also Table \ref{tab:keybaofits} for additional BAO fitting details.} 
Moreover, the errors tend to improve significantly after reconstruction for about 90\% of the cases (i.e., right panel).
Hence, the reconstruction procedure appears to be efficient on the mocks. 
 
Reconstruction applied to LRG data appears instead to 
produce only marginal effects.
To this end, in Figure \ref{fig:scatter_pre-post_LRG}
we overplot the \desimtwo\ LRG measurements (cyan stars in the left and right panels), 
obtained by using the 
\texttt{RascalC}-LRG covariance calibrated directly on LRGs.  
As evident from the figure, 
the \desimtwo\ LRG measurements
are located in the upper right corner of the significance scatter plot (left panel),
and in the lower left corner of the $\sigma_{\alpha}$ plot (right panel), respectively:
hence,  we are in a similarly `lucky' situation as those reported for the
BOSS CMASS LRG sample by \citet{BOSSDR9BAO}, 
and also for eBOSS LRGs \citep{Bautista2021}.
Table \ref{tab:keybaofits}  provides a quantification of the `lucky' realization of the observational 
data point, showing that
both its detection significance and precision are consistent
with those obtained via mock averages. In particular, 
 focusing on post-reconstruction results, the detection significance of the data 
point is 5.050 (in units of standard deviations) with a precision of 1.7\%, 
while the average \texttt{EZmocks} results yield a detection significance of 4.138 with a precision of 2.1\%,
and from the average of the \texttt{AbacusSummit} mocks we obtain
4.242 with a 2.0\% precision.
Note that  a small $\sigma_{\alpha}$ implies a better BAO detection, thus
a higher significance. In essence, while
generally reconstruction improves errors on $\alpha$,
this may not happen if the starting (pre-reconstruction) point already has a low error to begin with (i.e., a `lucky' realization).
In such a situation, reconstruction does not tend to produce much improvement, as shown in the $\sim 10\%$ of the mocks in our analysis.
This seems to be the case  for the \desimtwo\ LRG sample data volume:
our recovered $\sigma_{\alpha}$ for data is much smaller than the mean expected
from the mocks (right panel), and our BAO detection significance is high (left panel),
showing a strong and well-defined
acoustic peak. 
 
%--------------------------------------------------------------- 
%--------------------------------------------------------------- 

\subsection{BAO Detection from the LRG Sample}

%-------------------------------%

\begin{table*}
\centering
\caption{BAO key fitting results for \desimtwo\ LRGs and BGS.}
\begin{tabular}{c|c|c|c|c}
 \hline\hline 
 Sample & Reconstruction & BAO Detection Significance & $\alpha + \Delta \alpha$  & min$(\chi^2)$/dof \\
 \hline
 \desimtwo\ LRG & Pre-recon & 5.170 & 0.987 $\pm$ 0.016 & 15.619 / 20\\
                & Post-recon & 5.050 & 1.000 $\pm$ 0.017 & 13.463 / 20 \\
 \hline
 \desimtwo\ BGS & Pre-recon &  2.337  & 0.980 $\pm$ 0.040 & 13.172 / 20\\
                & Post-recon & 2.963 & 1.001  $\pm$ 0.026 & 16.724 / 20\\
 \hline\hline
\end{tabular}
\label{tab:keybaofits_data}
\end{table*}

%-------------------------------% 

\begin{figure*}
\centering
\includegraphics[width=0.49\textwidth]{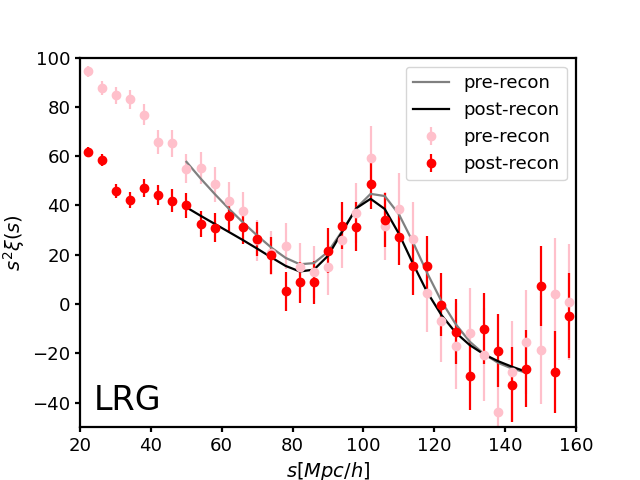}
\includegraphics[width=0.49\textwidth]{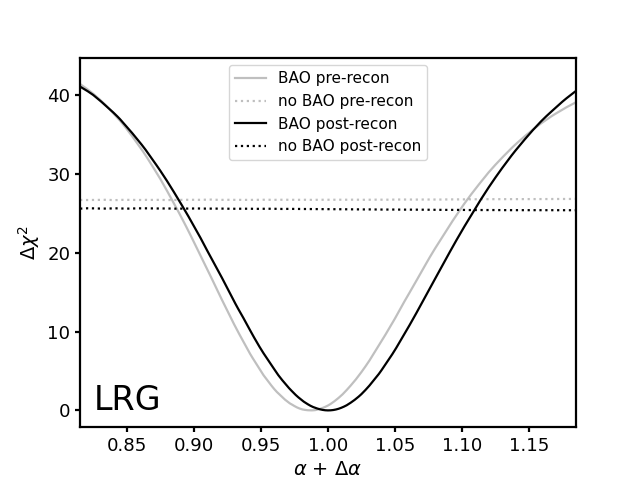}
\caption{BAO feature and its significance, as detected in the large-scale correlation function of \desimtwo\ LRGs. 
[Left] Pre- (lighter pink dots with errorbars) and post- (red dots with errorbars) reconstruction two-point clustering 
statistics inferred from the LRG sample, clearly
displaying the BAO peak.  The gray and black lines are respectively pre- and post-reconstruction fits to
$\xi(s)$ in the spatial range $50-150h^{-1}{\rm Mpc}$ over 25 points with 20 dof, obtained using the 
\texttt{RascalC-LRG} covariance. 
Errorbars are the square root of its diagonal elements. 
[Right] Detection significance of the  \desimtwo\ LRG BAO feature before (lighter gray lines)
and after (black lines) reconstruction. Dotted lines with 
similar colors are corresponding fits to the data using a model without BAO. 
We have shifted each value of $\alpha$ by $\Delta\alpha$ both in pre- and post-reconstruction, 
such that the minimum $\chi^2$ of the post-reconstruction result is at 1.
We note that an identical $\Delta\alpha$ was introduced for post-reconstruction LRGs and BGS 
(Figure \ref{fig:det-sign_post_BGS}), 
to demonstrate the coherence in terms of cosmological implications 
from the two tracers at the two different redshifts, 
while being blinded.
The BAO peak is detected
at $\sim 5\sigma$ confidence with a $1.6 \%$ and $1.7 \%$ precision 
in the pre- and post-reconstruction \desimtwo\ LRG sample, respectively, 
with the reconstruction procedure playing only a marginal role.  
Such a remarkable detection level, obtained with just two months
of DESI operations, is comparable to the one reported for the
BOSS CMASS sample \citep{BOSSDR9BAO}, and it is quite reassuring -- given the 
high complexity of the DESI instrument and of the DESI spectroscopic reduction pipeline.}
\label{fig:det-sign_post_LRG}
\end{figure*}

%-------------------------------% 

\begin{figure*}
\centering
\includegraphics[width=0.49\textwidth]{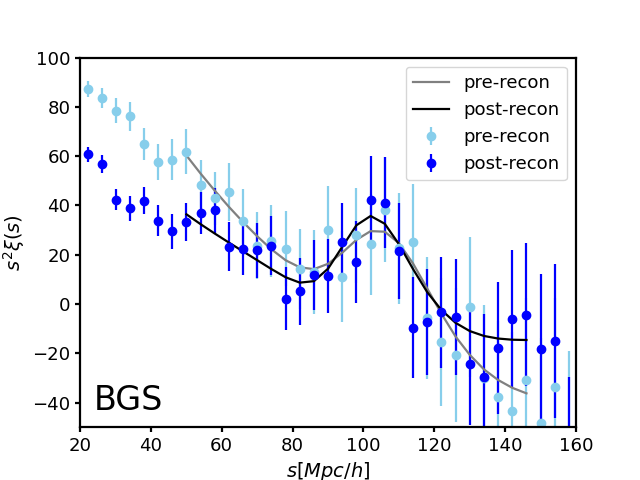}
\includegraphics[width=0.49\textwidth]{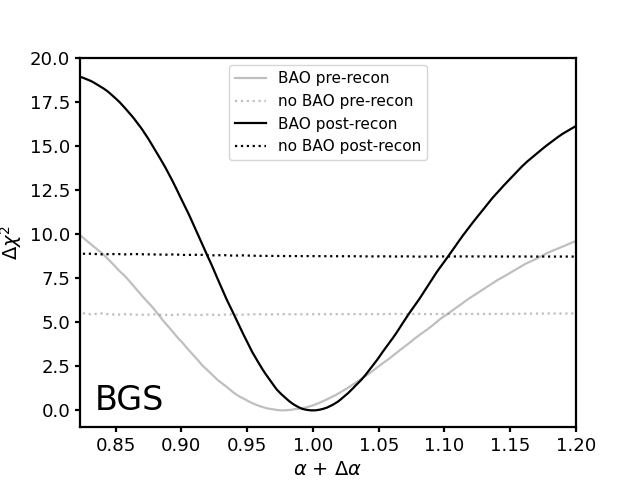}
\caption{BAO feature and its significance, as detected in the large-scale correlation function of \desimtwo\ BGS.  
Line styles and conventions same as in Figure \ref{fig:det-sign_post_LRG}. 
[Left] Pre-  and post-reconstruction two-point clustering statistics inferred from the BGS sample.
BAO fits to $\xi(s)$ are performed in the spatial range $50-150h^{-1}{\rm Mpc}$,  adopting the
\texttt{RascalC-BGS} covariance. Errorbars are the square root of its diagonal elements. 
[Right] Detection significance of the  \desimtwo\ BGS BAO feature before (lighter gray lines)
and after (black lines) reconstruction. Dotted lines with 
indentical colors are corresponding fits to the data using a model without BAO. 
We have shifted each value of $\alpha$ by the
same $\Delta\alpha$ both
in pre- and post-reconstruction, as done for the LRG analysis. 
The acoustic feature is detected at $\sim 2.5\sigma$ significance with a  $4.0\%$ precision in pre-reconstruction,
and at $\sim 3.0\sigma$ with a  $2.6 \%$ precision in post-reconstruction. Clearly, for this galaxy sample,
reconstruction plays a more substantial effect in sharpening the acoustic peak.
This BAO detection represents another relevant milestone of our  \desimtwo\ analysis.}
\label{fig:det-sign_post_BGS}
\end{figure*}
 
%-------------------------------% 

Figure  \ref{fig:det-sign_post_LRG}
displays the BAO fit to the \desimtwo\ LRG two-point correlation function, 
along with its significance:
this measurement represents one of the key results of our analysis. 
The left panel shows the pre- and post-reconstruction clustering statistics computed with the 
LS estimator (points with errorbars), and the best-fit model (curves). 
The gray and black curves are respectively pre- and post-reconstruction fits to
$\xi(s)$ in the spatial range $50-150h^{-1}{\rm Mpc}$,\footnote{Since we assume a bin size of
$4h^{-1}{\rm Mpc}$ for characterizing the $\xi(s)$ clustering statistics, we therefore
fit over 25 points using five parameters, leaving us 20 degrees-of-freedom (dof).} obtained with the procedure detailed in Section \ref{subsec:methods:fitting} and using the \texttt{RascalC}-LRG covariance matrix:  
errorbars in the plot show the square root of its diagonal elements. 
The BAO peak is clearly detected, and well matched to the best-fitting model. This is confirmed quantitatively: we find $\chi^2_{\rm min} =15.6$ and $\chi^2_{\rm min} =13.5$ for the pre- and post-reconstruction cases assuming
20 dof, respectively. Table \ref{tab:keybaofits_data} reports the specifics of these BAO fits.

The right panel of Figure \ref{fig:det-sign_post_LRG} displays the likelihood for the \desimtwo\ LRG BAO scale, 
as represented by $\Delta \chi^2 = \chi^2 - \chi^2_{\rm min}$, before and after reconstruction (solid curves). 
Dotted lines having identical colors represent corresponding fits to the data using
a model without BAO.
This provides two crucial results: the uncertainty on the measurement, and the significance of the BAO feature. 
Assuming a Gaussian likelihood, the 1$\sigma$ confidence region is represented by the width of the curve with $\Delta \chi^2 < 1$. We estimate the 1$\sigma$ uncertainty to be 0.016 in pre-reconstruction and 0.017 in post-reconstruction, respectively. 
The BAO detection significance can be simply determined by comparing results obtained from a fit to the data using a model without BAO (displayed via dotted lines in the figure), and once again subtracting the $\chi^2_{\rm min}$ from the BAO fit. 
This indicates how much better a model containing BAO fits the LRG data (i.e., actual existence of the BAO peak in the galaxy sample). The $\Delta \chi^2_{\rm min, no BAO}$ is greater than 25 both in pre- and post-reconstruction. 
Hence, we report a detection of the BAO feature in the \desimtwo\ LRG sample at a significance greater than 5$\sigma$.

We note that  we have shifted each $\alpha$ by the corresponding value of $\Delta \alpha$ 
in the right panel of Figure \ref{fig:det-sign_post_LRG},
such that the minimum $\chi^2$ is at 1 for the post-reconstruction case. 
The magnitude of the required shift for the post-reconstruction result was less than 1$\sigma$. 
Thus, while we do not reveal the precise value of the BAO scale in this analysis, we are 
consistent with the fiducial cosmology. 
 
As illustrated in  Figure  \ref{fig:det-sign_post_LRG}, the
BAO detection in the   \desimtwo\ LRG sample is highly significant.
Such a remarkable detection level, obtained with only two months
of DESI operations, is comparable to the one reported for the
BOSS high-$z$ LRG sample \citep[i.e., CMASS;][]{BOSSDR9BAO},
comprised of 264 283 galaxies in the redshift interval $0.43 < z < 0.7$.  
Notice also that the reconstruction procedure has practically 
no impact on the BAO peak inferred from the \desimtwo\ LRG clustering,
as evident from the right panel of Figure  \ref{fig:det-sign_post_LRG} 
(compare the gray and black curves). 
As pointed out in the previous section, this is
due to the `lucky' starting point of the
pre-reconstruction LRG measurement,  
which happens to be located in the
lower left corner of the $\sigma_{\alpha}$ plot in Figure \ref{fig:scatter_pre-post_LRG}, 
yielding already a very low error to begin with, and 
hence carrying a high BAO detection significance 
(i.e., left panel of the same figure, see the cyan star 
in the upper right corner of the significance scatter plot).
 
In closing this section, we emphasize that the first BAO 
measurement obtained with \desimtwo\ LRGs 
represents an  important milestone, and its high detection 
level is quite reassuring -- considering the 
complexity of the DESI instrument and of the spectroscopic reduction pipeline.  
It also constitutes an important early validation and 
quality-control of the data management system, as well as
a confirmation of the successful survey design strategy adopted for DESI targets. 
Next, we move to the BGS sample and 
carry out a similar BAO analysis.

%--------------------------------------------------------------- 
%--------------------------------------------------------------- 

\subsection{BAO Detection from the BGS}

%-------------------------------% 

Figure \ref{fig:det-sign_post_BGS} contains another central result of our analysis.
Here, we report the BAO feature as detected in the large-scale
clustering of the \desimtwo\ BGS BRIGHT sample characterized by a magnitude cut 
of $-21.5$ (see Section \ref{sec:datasamples}),  
together with its significance. In detail, 
the left panel displays two-point  
correlation function measurements from those galaxies. 
Following similar conventions as in Figure \ref{fig:det-sign_post_LRG},
the gray and black curves are respectively the pre- and post-reconstruction best-fit models  to 
$\xi(s)$ in the spatial range $50-150h^{-1}{\rm Mpc}$,
obtained using the \texttt{RascalC-BGS} covariance matrix.
Errorbars in the plot show
the square root of its diagonal elements.
Also in this case, the 
BAO peak is clearly detected and
well matched to the best-fitting model.
Specifically, $\chi^2_{\rm min} =13.2/20~{\rm dof}$ in pre-reconstruction,
and $\chi^2_{\rm min} =16.7/20~{\rm dof}$ in post-reconstruction.
See again Table \ref{tab:keybaofits_data} for details on these BAO fits.

The right panel of Figure \ref{fig:det-sign_post_BGS} 
shows the likelihood for the \desimtwo\ 
BGS BRIGHT BAO scale before and after reconstruction (solid lines).  
Dotted lines with identical colors represent the 
corresponding fits to the data using a model without BAO.
Similarly to the LRG analysis, 
we determine the 1$\sigma$ confidence region of the measurements based on the 
width of the curve with $\Delta \chi^2 < 1$. This yields an uncertainty 
of $0.040$ in pre-reconstruction, and of $0.026$ in post-reconstruction. 
As evident by comparing the gray and black curves
from both panels of Figure \ref{fig:det-sign_post_BGS},
here reconstruction plays a more substantial effect in 
sharpening the acoustic peak
and in partially removing the BAO smearing caused by non-linear 
structure growth. 
By comparing the $\chi^2$ of the data fits against a 
model without BAO, we determine the significance of the BAO feature. 
We find $\Delta \chi^2$ to be 5.5 in pre-reconstruction and 8.8 
in post-reconstruction, corresponding to a 
BAO detection significance
of $\sim 2.3 \sigma$ and $\sim 3.0 \sigma$, respectively.  
 
Even for the \desimtwo\ BGS sample, we
have shifted the individual $\alpha$'s by the 
same $\Delta \alpha$ factor applied to the LRG sample,
both in pre- and post-reconstruction. 
Since the BGS minimum $\chi^2$ values are within $1\sigma$ of $\alpha + \Delta \alpha =1$, 
we can conclude that the BGS BAO results are fully consistent with the LRG ones.

Along with the first BAO measurement from the \desimtwo\ LRG sample,
this first BAO detection obtained using  the \desimtwo\ BGS represents 
another relevant milestone, as well as an additional  early validation of the 
DESI pipeline and data management system for the bright time survey.
  
%%%%%%%%%%%%%%%%%%%%%%%%%%%%%%%%%%%%%%%%%%%%%%%%%%
%%%%%%%%%%%%%%%%%%%%%%%%%%%%%%%%%%%%%%%%%%%%%%%%%%
%%%  FUTURE

\section{Outlook for Future Data Releases}\label{sec:outlook}

%--------------------------------------------------------------- 
%--------------------------------------------------------------- 

 Based on the promising BAO results
presented in Section \ref{sec:results}, 
obtained with the \desimtwo\ dataset collected over just two months of operations, 
we now proceed to forecast the expected 
BAO detection significance and accuracy 
with the completed survey data, 
focusing on the final Y5 DESI LRG sample.

To this end, we compute 
Fisher matrix forecasts of the LRG isotropic BAO scale.
We perform such calculations by  
adopting  covariance matrices constructed using post-reconstruction 
\texttt{EZmocks} for the two survey configurations (i.e., \desimtwo\ LRGs and DESI Y5 LRGs, respectively)
and the best-fit monopole model for \desimtwo\ -- while marginalizing over the same free parameters as 
in the actual data fit. We then 
take the ratio of the two Fisher estimates to rescale the \desimtwo\ constraint,
in order to account for the specific calibration
of the LRG \texttt{EZmocks}, tuned on the One Percent Survey clustering 
rather than on \desimtwo\  (see again Section \ref{subsec:calibration} for details). 
These Fisher estimates return a factor of $\sim 5.8$ between the BAO SN
inferred from Y5 LRGs and that of \desimtwo\ LRGs. 
When we rescale our data best-fit LRG estimate of $1.7\%$ accounting for this 
factor, we then predict a $\sim 0.29\%$ precision 
on the BAO scale from the final Y5 LRG sample over $0.4<z<1.1$. 

We note that such estimate  should be taken as an approximate projection for DESI Y5, since we have simply assumed the same 
BAO signal from \desimtwo\ while only changing the covariance. Nevertheless, 
the projected BAO precision for DESI Y5 LRGs agrees well with the more accurate 
DESI LRG Y5 forecasts based on {\tt GoFish}\footnote{https://github.com/ladosamushia/GoFish}
-- i.e., 0.25\% precision for the DESI Y5 LRG sample, presented in \cite{DESIsv}. 
This exquisite level of sub-percent precision on the BAO scale (even from a single tracer)
will confirm DESI as the most competitive BAO experiment for the remainder of this decade. 

%%%%%%%%%%%%%%%%%%%%%%%%%%%%%%%%%%%%%%%%%%%%%%%%%%
%%%%%%%%%%%%%%%%%%%%%%%%%%%%%%%%%%%%%%%%%%%%%%%%%%
%%%  CONCLUSION

\section{Conclusions}\label{sec:conclusion}

%--------------------------------------------------------------- 
%--------------------------------------------------------------- 

 The BAO scale represents a key standard ruler that provides a direct way to measure the expansion history of the Universe 
and infer robust cosmological constraints. 
Hence, BAO measurements are considered primary DESI science targets, and a major deliverable at any stage of the survey. 
Precision on the expansion history of the Universe constitutes a compelling
probe of the nature of DE. DESI is expected to deeply impact the current 
understanding of DE, 
along with providing unprecedented constrain on theories of modified gravity and inflation, 
and on neutrino masses \citep{DESICollaboration2016a}.
In this respect, DESI plans to conduct a series of BAO analyses throughout its five-year survey time with blinded catalogs: 
the Y1 sample would be the first of such rigorously blinded BAO analyses. 

The remarkable complexity of the DESI instrument, along with the adopted survey design
and the elaborated DESI spectroscopic pipeline and data management system, pose a  
potential challenge to all BAO analyses. It is then of utmost importance to test  such pipelines well in advance, and 
it has been the main goal of the current study:
this is a crucial aspect in order to guarantee the success of future BAO investigations, and
for confirming the optimal performance of the DESI spectrograph and the quality level of the data reduction pipeline.
Precisely for this reason, the 
first two months of DESI operations were intentionally 
kept unblinded (i.e., \desimtwo\ sample).\footnote{Note that \desimtwo\ is 
approximately 1/5 of the DESI Y1 sample.}  
 
To this end, we have used the  \desimtwo\ dataset
and reported the first high-significance detection of the BAO
signal from the LRG and BGS samples. 
Specifically, our primary results are:
\begin{itemize}
\item $\sim 5\sigma$ level BAO detection in the \desimtwo\ LRG sample at a precision of $1.7\%$.
\item $\sim 3\sigma$ level BAO detection in the \desimtwo\  BGS sample at a precision of $2.6\%$.  
\end{itemize}
In particular, our LRG BAO measurement is comparable to the  
$5-6 \sigma$ BAO detection obtained with the BOSS high-$z$ LRG sample 
\citep[i.e., CMASS;][]{BOSSDR9BAO}, 
comprised of 264,283 galaxies in 
the redshift interval $0.43 < z < 0.7$. 
Moreover, the BOSS and eBOSS BAO measurements made with LRGs 
between $0.4<z<1.0$ (with $N_{\rm gal}= 1{,}063{,}828$) returned an 
aggregate precision of 0.77\% on $D_{\rm V}$ \citep{Bautista2021}, 
which is only a factor of $2.2$ times better (in terms of precision) than our 
quoted result with \desimtwo\ LRGs  (having just $N_{\rm gal}= 266,269$).
This latter aspect is rather remarkable, considering that the
\desimtwo\ dataset has been collected simply
during the initial two months of DESI operations. 

Based on these results, we
forecasted that DESI is
on target to achieve a
high-significance BAO detection 
at  a  $\sim 0.29\%$ precision
with the completed
Y5 LRG sample over $0.4<z<1.1$, 
meeting the DESI
top-level science requirements on BAO measurements.  
This exquisite level of precision 
will set novel standards in cosmology and
confirm DESI as a highly accurate and precise Stage-IV BAO experiment.

Additional relevant aspects of our investigation on these preliminary \desimtwo\ data
that are worth highlighting are summarized as follows: 
\begin{itemize}
\item Although the catalogs we used are unblinded, we presented a blinded 
cosmology analysis -- in that we do not report here the best fit BAO scale.  
In fact, we only presented the precision and detection level of the BAO measurements. We plan to provide a full cosmological 
interpretation with the Y1 data release in the near future, after additional rigorous systematic tests.  
\item We focused on the isotropic BAO scale exploiting only the monopole of the LRG and BGS samples. 
\item We applied the nominal BAO pipeline that has been previously well-tested with BOSS and eBOSS data. In particular, 
we utilized the early version of the pipeline 
package {\tt cosmodesi} that the DESI collaboration team has been developing, which wraps both existing and new cosmological 
galaxy survey analysis pipelines from the literature. 
\item We constructed, calibrated, and used semi-analytical semi-empirical covariance matrices 
based on the \texttt{RascalC} code, and validated those covariances in terms of 
BAO fitting procedures (in pre- and post-reconstruction)
using a set of mocks -- as detailed in Section \ref{sec:cov} and in \cite{RascalC-DA02}.
\item We also found that 
the LRG BAO signal from the \desimtwo\ data is   
stronger than the typical BAO signal present in the LRG mocks. 
Partly for this reason, the reconstruction procedure performed on actual LRG data is 
less effective than the one performed on LRG mocks. 
On the other hand, we found that the BGS sample shows a factor of $\sim 1.5$ precision 
improvement after reconstruction.
These results are consistent with the typical behavior we find on \desimtwo\ mocks; less scatter is expected for the more complete DESI Y1 and Y5 samples.
\end{itemize}

This work represents the first step towards the analysis techniques 
that will lead to the key cosmological results from DESI Y1 data.
While the BAO results presented here constitute an  important milestone
and are quite reassuring in terms of consistency in the clustering amplitude (considering the 
complexity of the DESI instrument and of the spectroscopic reduction pipeline), 
we anticipate that the DESI Y1 analysis alone will surpass the cosmological information from all of 
the BAO analyses performed to date. 
This will require going beyond the legacy BAO analysis setting that has been 
well-tested using BOSS and eBOSS data (and also mainly adopted here), with 
an unprecedented level of BAO systematic tests and by developing an optimal 
BAO pipeline -- given the stringent requirements on 
theoretical and observational systematics that are imposed by a dataset as powerful 
as we expect by the end of the survey.  
The DESI team is currently working in this direction, 
and presenting all these novel technical aspects
will be the subject of many forthcoming DESI Y1 cosmology papers.

%%%%%%%%%%%%%%%%%%%%%%%%%%%%%%%%%%%%%%%%%%%%%%%%%%
%%%%%%%%%%%%%%%%%%%%%%%%%%%%%%%%%%%%%%%%%%%%%%%%%%
%%%  ACKNOWLEDGEMENTS

\section*{Acknowledgements}

JM and GR acknowledge support from   
the National Research Foundation of Korea (NRF) through grant No. 2020R1A2C1005655 funded by 
the Korean Ministry of Education, Science and Technology (MoEST) and from  
the faculty research fund of Sejong University in 2022/2023. 
DV acknowledges support from the U.S. Department of Energy, Office of Science, 
Office of High Energy Physics under grant No. DE-SC0019091 and No. DE-SC0023241.
MR is supported by U.S. Department of Energy grant DE-SC0007881 and by the Simons Foundation Investigator program.
CS acknowledges support from the National Research Foundation of Korea (NRF) through grant 
No. 2021R1A2C101302413 funded by the Korean Ministry of Education, Science and Technology (MoEST). 
CS was also supported by the high performance computing cluster Seondeok at the Korea 
Astronomy and Space Science Institute. 

This research is supported by the Director, Office of Science, Office of High Energy Physics of the U.S. Department of Energy under Contract No. DE–AC02–05CH11231, and by the National Energy Research Scientific Computing Center, a DOE Office of Science User Facility under the same contract; additional support for DESI is provided by the U.S. National Science Foundation, Division of Astronomical Sciences under Contract No. AST-0950945 to the NSF’s National Optical-Infrared Astronomy Research Laboratory; the Science and Technologies Facilities Council of the United Kingdom; the Gordon and Betty Moore Foundation; the Heising-Simons Foundation; the French Alternative Energies and Atomic Energy Commission (CEA); the National Council of Science and Technology of Mexico (CONACYT); the Ministry of Science and Innovation of Spain (MICINN), and by the DESI Member Institutions: \url{https://www.desi.lbl.gov/collaborating-institutions}.

This work was led under the supervision of the DESI Y1 catalog generation Key Project conveners 
(Arnaud de Mattia and Ashley Ross) and the DESI Y1 BAO Key Project conveners (Nikhil Padmanabhan and Hee-Jong Seo).

We thank Antonella Palmese, Eric Linder, and Benjamin Weaver
for constructive comments and useful feedback. 

The authors are honored to be permitted to conduct scientific research on Iolkam Du'ag (Kitt Peak), 
a mountain with particular significance to the Tohono O'odham Nation.

%%%%%%%%%%%%%%%%%%%%%%%%%%%%%%%%%%%%%%%%%%%%%%%%%%
%%%%%%%%%%%%%%%%%%%%%%%%%%%%%%%%%%%%%%%%%%%%%%%%%%
%%%  DATA AVAILABILITY

\section*{Data Availability}
 
The redshift measurements used in this paper -- derived from the DESI \texttt{Guadalupe} dataset -- 
will be made public with the DESI Y1 data release (DR1). The analogous spectroscopic data reductions and 
redshift fits for the DESI EDR (i.e., \texttt{Fuji}) will be publicly available on NERSC at this url: \url{https://data.desi.lbl.gov/public/edr/spectro/redux/fuji}.
All data points shown in the published graphs are available in machine-readable 
form in Zenodo at \url{https://doi.org/10.5281/zenodo.7835433}.

%%%%%%%%%%%%%%%%%%%%%%%%%%%%%%%%%%%%%%%%%%%%%%%%%%
%%%%%%%%%%%%%%%%%%%%%%%%%%%%%%%%%%%%%%%%%%%%%%%%%%
%%%  REFERENCES

\bibliographystyle{mnras}
\bibliography{references}  

%%%%%%%%%%%%%%%%%%%%%%%%%%%%%%%%%%%%%%%%%%%%%%%%%%
%%%%%%%%%%%%%%%%%%%%%%%%%%%%%%%%%%%%%%%%%%%%%%%%%%
%%%  APPENDIX

\appendix
\section{Supporting Material} \label{sec:appendix} 

%--------------------------------------------------------------- 
%--------------------------------------------------------------- 

%-------------------------------%

\begin{table*}
\centering
\caption{BAO key fitting results for LRGs. For \desimtwo\ LRGs, we use the {\tt RascalC}-LRG covariance, 
while for the two sets of mocks we applied instead the 
{\tt RascalC}-EZ covariance -- see Section \ref{sec:cov} and Table \ref{tab:covariance} for details. 
We do not show here the $\alpha$ values for the LRG data fits, as the cosmology is kept blinded. 
Moreover, we do not report \desimtwo\ fits with the \texttt{EZmock} numerical covariance, since
\texttt{EZmocks} are calibrated with the One Percent Survey data 
(as explained in Section \ref{subsec:calibration}).}
\begin{tabular}{c|c|c|c|c}
 \hline\hline 
 & Reconstruction & BAO results & {\tt RascalC} cov & {\tt EZmock} cov \\
 \hline
 & Pre-recon & Detection significance & 5.170 & -- \\
 \desimtwo\ LRG & & Precision & 1.6\% & -- \\
 & Post-recon & Detection significance & 5.050 & -- \\
 & & Precision & 1.7\% & -- \\
 \hline
 & & Detection significance & 3.423 & 3.597 \\
 & Pre-recon & $\alpha$ & 1.006 & 1.006 \\
 EZmock LRG & & Precision & 2.8\% & 2.7\% \\
 & & Detection significance & 4.138 & 4.091 \\
 & Post-recon & $\alpha$ & 1.000 & 1.001 \\
 & & Precision & 2.1\% & 2.3\% \\
 \hline
 & & Detection significance & 3.623 & 3.801 \\
 & Pre-recon & $\alpha$ & 1.001 & 0.997 \\
 AbSmock LRG& & Precision & 2.8\% & 2.5\% \\
 & & Detection significance & 4.242 & 4.209 \\
 & Post-recon & $\alpha$ & 0.992 & 0.994\\
 & & Precision & 2.0\% & 2.1\% \\
 \hline\hline
\end{tabular}
\label{tab:keybaofits}
\end{table*}

%-------------------------------%

In support of the primary BAO analysis 
carried out in the main text, 
in Table \ref{tab:keybaofits} we provide some further technical details 
related to the various BAO fits performed to the \desimtwo\ LRG sample, as well as
to the corresponding LRG mocks adopted in this work -- namely, pre- an post-reconstruction results 
of the BAO detection significance and precision, along with the $\alpha$ values for the mock fits. 
Specifically, as reported in the main text,
the  \desimtwo\ LRG sample provides a 
$\sim5 \sigma$ BAO detection significance
at a 1.6\% and 1.7\% precision in pre- and post-reconstruction, respectively.
From the two sets of mocks considered (\texttt{AbacusSummit} and \texttt{EZmocks}), we
also find a significant BAO detection at more than  3.4 $\sigma$ in pre-reconstruction,
and  exceeding a 4.0 $\sigma$ detection in post-reconstruction, with a 
corresponding precision better than 2.8\% (pre-reconstruction)
or 2.3\% (post-reconstruction). Remarkably, 
the $\alpha$-values inferred from the mocks are close to unity,
indicating that the fiducial cosmology is very-well recovered. 
This also implies that the mock production pipeline is working properly.
Additionally, from Table \ref{tab:keybaofits} one can readily infer that 
the {\tt RascalC}-based LRG covariance is compatible with the \texttt{EZmock} LRG covariance
(as we also reported in Section \ref{sec:cov}),  
simply by comparing all the corresponding fitting results 
obtained with the two sets of covariances (i.e., last two columns). 

%%%%%%%%%%%%%%%%%%%%%%%%%%%%%%%%%%%%%%%%%%%%%%%%%%
%%%%%%%%%%%%%%%%%%%%%%%%%%%%%%%%%%%%%%%%%%%%%%%%%%
%%%  AFFILIATIONS

\section*{} 
 
%--------------------------------------------------------------- 
%--------------------------------------------------------------- 

% List of Institutions
$^{1}$Department of Physics and Astronomy, Sejong University, Seoul, 143-747, Korea\\
$^{2}$Department of Physics \& Astronomy, Ohio University, Athens, OH 45701, USA \\
$^{3}$Center for Astrophysics $|$ Harvard \& Smithsonian, 60 Garden Street, Cambridge, MA 02138, USA\\
$^{4}$Korea Astronomy and Space Science Institute, 776 Daedeok-daero, Yuseong-gu, Daejeon 34055, Republic of Korea\\
$^{5}$Lawrence Berkeley National Laboratory, 1 Cyclotron Road, Berkeley, CA 94720, USA\\
$^{6}$Physics Dept., Boston University, 590 Commonwealth Avenue, Boston, MA 02215, USA\\
$^{7}$Tata Institute of Fundamental Research, Homi Bhabha Road, Mumbai 400005, India\\
$^{8}$Physics Department, Yale University, P.O. Box 208120, New Haven, CT 06511, USA\\
$^{9}$NSF's NOIRLab, 950 N. Cherry Ave., Tucson, AZ 85719, USA\\
$^{10}$Department of Physics \& Astronomy, University College London, Gower Street, London, WC1E 6BT, UK\\
$^{11}$IRFU, CEA, Universit\'{e} Paris-Saclay, F-91191 Gif-sur-Yvette, France\\
$^{12}$Department of Physics and Astronomy, The University of Utah, 115 South 1400 East, Salt Lake City, UT 84112, USA\\
$^{13}$Instituto de F\'{\i}sica, Universidad Nacional Aut\'{o}noma de M\'{e}xico, Cd. de M\'{e}xico C.P. 04510, M\'{e}xico\\
$^{14}$Department of Physics, Southern Methodist University, 3215 Daniel Avenue, Dallas, TX 75275, USA\\
$^{15}$Fermi National Accelerator Laboratory, PO Box 500, Batavia, IL 60510, USA\\
$^{16}$Institut de F\'{i}sica dAltes Energies (IFAE), The Barcelona Institute of Science and Technology, Campus UAB, 08193 Bellaterra Barcelona, Spain\\
$^{17}$Departamento de F\'isica, Universidad de los Andes, Cra. 1 No. 18A-10, Edificio Ip, CP 111711, Bogot\'a, Colombia\\
$^{18}$Department of Physics, The University of Texas at Dallas, Richardson, TX 75080, USA\\
$^{19}$University of Michigan, Ann Arbor, MI 48109, USA\\
$^{20}$Department of Physics, The Ohio State University, 191 West Woodruff Avenue, Columbus, OH 43210, USA\\
$^{21}$Center for Cosmology and AstroParticle Physics, The Ohio State University, 191 West Woodruff Avenue, Columbus, OH 43210, USA\\
$^{22}$Department of Physics, Southern Methodist University, 3215 Daniel Avenue, Dallas, TX 75275, USA\\
$^{23}$Natural Science Research Institute, University of Seoul, 163 Seoulsiripdae-ro, Dongdaemun-gu, Seoul, South Korea\\
$^{24}$Sorbonne Universit\'{e}, CNRS/IN2P3, Laboratoire de Physique Nucl\'{e}aire et de Hautes Energies (LPNHE), FR-75005 Paris, France\\
$^{25}$Departament de F\'{i}sica, Serra H\'{u}nter, Universitat Aut\`{o}noma de Barcelona, 08193 Bellaterra (Barcelona), Spain \\
$^{26}$Department of Astronomy, The Ohio State University, 4055 McPherson Laboratory, 140 W 18th Avenue, Columbus, OH 43210, USA\\
$^{27}$Instituci\'{o} Catalana de Recerca i Estudis Avan\c{c}ats, Passeig de Llu\'{\i}s Companys, 23, 08010 Barcelona, Spain\\
$^{28}$Department of Physics and Astronomy, Siena College, 515 Loudon Road, Loudonville, NY 12211, USA\\
$^{29}$Department of Physics \& Astronomy, University of Wyoming, 1000 E. University, Dept.~3905, Laramie, WY 82071, USA\\
$^{30}$Institute of Cosmology \& Gravitation, University of Portsmouth, Dennis Sciama Building, Portsmouth, PO1 3FX, UK\\
$^{31}$Institute for Astronomy, University of Edinburgh, Royal Observatory, Blackford Hill, Edinburgh EH9 3HJ, UK\\
$^{32}$Department of Physics \& Astronomy and Pittsburgh Particle Physics, Astrophysics, and Cosmology Center (PITT PACC), University of Pittsburgh, 3941 O'Hara Street, Pittsburgh, PA 15260, USA\\
$^{33}$National Astronomical Observatories, Chinese Academy of Sciences, A20 Datun Rd., Chaoyang District, Beijing, 100012, P.R. China\\
$^{34}$Waterloo Centre for Astrophysics, University of Waterloo, 200 University Ave W, Waterloo, ON N2L 3G1, Canada\\
$^{35}$Perimeter Institute for Theoretical Physics, 31 Caroline St. North, Waterloo, ON N2L 2Y5, Canada\\
$^{36}$Space Sciences Laboratory, University of California, Berkeley, 7 Gauss Way, Berkeley, CA 94720, USA\\ 
$^{37}$GMT RPG - Campus de Excelencia Internacional CIE/UAM + CSIC, Universidad Autónoma de Madrid\\
$^{38}$Department of Physics, Kansas State University, 116 Cardwell Hall, Manhattan, KS 66506, USA \\
$^{39}$Ecole Polytechnique F\'{e}d\'{e}rale de Lausanne, CH-1015 Lausanne, Switzerland\\
$^{40}$Department of Physics, University of California, Berkeley, 366 LeConte Hall MC 7300, Berkeley, CA 94720-7300, USA\\
$^{41}$SLAC National Accelerator Laboratory, Menlo Park, CA 94305, USA\\

%%%%%%%%%%%%%%%%%%%%%%%%%%%%%%%%%%%%%%%%%%%%%%%%%%
%%%%%%%%%%%%%%%%%%%%%%%%%%%%%%%%%%%%%%%%%%%%%%%%%%
%%%  END

\bsp 
\label{lastpage}
\end{document}